\documentclass[prd,nofootinbib,showpacs]{revtex4}

\renewcommand{\b}[1]{\bar{#1}}
\newcommand{\be}{\begin{equation}}
\newcommand{\ee}{\end{equation}}
\newcommand{\bea}{\begin{eqnarray}}
\newcommand{\eea}{\end{eqnarray}}

\newcommand{\Del}{\nabla}
\newcommand{\del}{\partial}

\newcommand{\N}{\mathcal{N}}
\renewcommand{\ap}{\alpha'}
\newcommand{\lst}{\ell_{\mathnormal{SM}}}
\newcommand{\lsh}{\ell_{h}}
\newcommand{\lss}{\ell_{\mathnormal{inf}}}

\newcommand{\comment}[1]{}
\newcommand{\msm}{M_{\mathnormal{SM}}}
\newcommand{\asm}{A_{\mathnormal{SM}}}
\newcommand{\ainf}{A_{\mathnormal{inf}}}

\begin{document}
\preprint{CALT-68-2570}
\preprint{NORDITA-2005-42}

\title{Stringy Effects During Inflation and Reheating}
\author{Andrew R. Frey}
\email{frey@theory.caltech.edu}
\affiliation{California Institute of Technology, 452-48\\
Pasadena, CA 91125, USA}

\author{Anupam Mazumdar}
\email{anupamm@nordita.dk}
\affiliation{NORDITA, Blegdamsvej -17\\ DK-2100, Copenhagen, Denmark}

\author{Robert Myers}
\email{rmyers@perimeterinstitute.ca}
\affiliation{Perimeter Institute for Theoretical Physics\\
 Waterloo, Ontario  N2L 2Y5, Canada\\ and\\
Department of Physics, University of Waterloo\\ Waterloo, Ontario
N2L 3G1, Canada}


\begin{abstract}
We consider inflationary cosmology in the context of string
compactifications with multiple throats. In scenarios where the
warping differs significantly between throats, string and Kaluza-Klein
physics can generate potentially observable corrections to the
cosmology of inflation and reheating.  First we demonstrate that a
very low string scale in the ground state compactification is
incompatible with a high Hubble scale during inflation, and we propose
that the compactification geometry is altered during inflation.  In
this configuration, the lowest scale is just above the Hubble scale,
which is compatible with effective field theory but still leads to
potentially observable CMB corrections. Also in the appropriate region
of parameter space, we find that reheating leads to a phase of long
open strings in the Standard Model sector (before the usual
radiation-dominated phase). We sketch the cosmology of the long string
phase and we discuss possible observational consequences.
\end{abstract}

\maketitle

\section{Introduction: Seeing Strings in Cosmology}\label{s:intro}

With the great success of inflationary cosmology in explaining the
spectrum of cosmic microwave background (CMB) fluctuations
\cite{Bennett:2003ba,Bennett:2003bz,Komatsu:2003fd,Peiris:2003ff}, as
well as recent progress in understanding inflation in string theory
--- see, for example,
\cite{Burgess:2001fx,Mazumdar:2001mm,Burgess:2001vr,
Garcia-Bellido:2001ky,Herdeiro:2001zb,Jones:2002cv,Kachru:2003sx,
Firouzjahi:2003zy,Burgess:2004kv,Hsu:2004hi,Dasgupta:2004dw,
Blanco-Pillado:2004ns,Becker:2005sg,Cremades:2005ir,Linde:2005dd,
Dimopoulos:2005ac,Cline:2005ty} --- the time is ripe to ask what
signals string theory might have in cosmology.  Historically
speaking, there has not been much cause for optimism.  If strings
(or other high energy physics, such as quantum gravity) modify
four dimensional effective field theory (EFT) above an energy
scale $M$, corrections to the usual Gaussian, scale-invariant CMB
spectrum enter at order $(H/M)^n$, with $n=2$ or 1. The value of
$n$ has been a matter of some debate in the literature. Standard
EFT arguments give $n=2$ \cite{Kaloper:2002uj} unless boundary
terms are added \cite{Burgess:2003zw,Schalm:2004qk,Porrati:2004gz,
Greene:2004np,Easther:2004vq,Collins:2005nu,Easther:2005yr,Collins:2005cm}.
The latter may yield $n\sim 1$ and seem to reproduce the effects
of an unconventional initial state or nonadiabatic evolution
\cite{Martin:2000xs,Brandenberger:2000wr,
Martin:2000bv,Danielsson:2002kx,Danielsson:2002qh,
Brandenberger:2002hs,Martin:2003kp,Brandenberger:2004kx,
Niemeyer:2000eh,Niemeyer:2001qe,Kempf:2001fa,Niemeyer:2002kh,Easther:2001fi,
Easther:2001fz,Easther:2002xe,Danielsson:2004xw,Burgess:2002ub}.
However, the main issue remains the conspicuously small value of
$H/M$. Experimental data appears to constrain
$H/M_{P}\lesssim 10^{-4}$. Even if the scale of
new physics is decoupled from the Planck scale, an optimistic
bound seems to be $H/M\lesssim 10^{-2}$ for realistic string
compactifications \cite{Kaloper:2002uj},
even including TeV gravity scenarios. 
Except in special cases, for
example \cite{Silverstein:2003hf,Alishahiha:2004eh}, string or
higher dimensional physics seems unobservable.  We will argue,
however, that a wide class of warped string compactifications
\textit{naturally} yields $H/M\sim 1$ during inflation, so that
stringy corrections to the EFT are potentially observable. In
addition, in the same models, thermal strings may dominate the
universe during reheating, leading to other potential cosmological
consequences.

In brief, our general argument runs as follows; the key ingredient is
warping in compactifications with more than one throat.  As explained
in \cite{Randall:1999ee}, warping of the spacetime dimensions
generates potentially large hierarchies of scales, possibly creating
sectors in which the string and Kaluza-Klein (KK) scales are low.
Significant warping can occur in string compactifications, including
those described in \cite{Giddings:2001yu,Kachru:2003aw}, and we will
consider models in which the standard model (SM) is confined to branes
in a highly warped region (the SM throat).  Since these
compactifications can also have all moduli frozen, they are suitable
both for low energy phenomenology and inflation
\cite{Kachru:2003sx}. However, we imagine that the inflationary
potential generated in some other region of the compactification (the
inflationary throat) than the SM throat. There is a tension between
the (usual) high scale of inflation and the low scale of fundamental
physics in the SM throat, however, since the 4D curvature $R\sim H^2$
can be much larger than the SM sector string scale.  It seems that the
low energy description of the SM throat must break down during
inflation.  We will give this general argument in section
\ref{s:naive}, first reviewing the specific compactifications and
inflationary models that we will consider as concrete examples.

The naive idea of inflation occurring on a fixed compactification
manifold, however, is in conflict with known physics from 4D EFT.
In section \ref{s:stringinflate}, we will describe a consistent
resolution of the tension between Hubble and SM scales.  Due to
cross-couplings and quantum fluctuations, inflation generates
Hubble-scale masses for all fields in EFT
\cite{Dvali:1995mj,Dine:1995uk,Dine:1995kz}.  From the 10D point
of view, the SM throat cannot support such massive modes unless
the minimum SM throat warp factor is roughly $H/M_P$, so we argue
that the compactification does not lie in its ground state during
inflation. Specifically, the SM throat is ``shorter'' than in the
post-inflationary ground state, and the SM sector string mass
scale is somewhat greater than the Hubble scale of inflation. We
end up with a consistent picture; with a large enough SM string
scale, EFT is a reasonable description of the physics after all.
As a bonus, the string and KK scales are naturally near the Hubble
scale, so high energy physics will more easily alter the
CMB.\footnote{In appendix \ref{s:alternatives}, we discuss
alternative possibilities for the SM throat geometry during
inflation and give reasons that they do not capture the
appropriate physics.}

After inflation, the question becomes how to reheat the SM degrees of
freedom and generate the usual hot big bang cosmology. In particular,
in the present scenario, the SM sector is physically separated from
the inflationary sector in the compactification
geometry. Concentrating on a specific inflationary model, we adopt the
argument of \cite{Barnaby:2004gg} that KK modes transmit energy
between the two sectors and give a more detailed estimate of the
reheating rate in section \ref{s:warpreheat}.\footnote{While this work
was being completed, a related discussion appeared in
\cite{Kofman:2005yz,Chialva:2005zy}.} 
We also note that the SM throat will relax to
its ground state during reheating, and, in fact, this relaxation will
itself reheat the SM sector as well. In the end, we find that, in a
large region of parameter space, either (or both) of these reheating
processes can yield a reheat temperature larger than the SM throat
string scale.  In that case, without deriving a string theoretic
reheating mechanism, we argue that reheating should lead to a phase of
long open strings on the SM branes at the Hagedorn temperature.  Some
additional comments on reheating, including sketches of reheating in
other inflationary models, are relegated to appendix
\ref{s:otherreheat}.

Following to the logical conclusion, we discuss the cosmological
evolution of open string matter on the SM branes in section
\ref{s:decayreheat}.  We first review the thermal distribution of open
strings from \cite{Lee:1997iz,Abel:1999rq,Barbon:2004dd} and argue
that, unlike closed strings, open strings can easily maintain thermal
equilibrium.  More details of string thermodynamics are given in
appendix \ref{s:morethermo}.  We then review the open string decays
that can reheat SM radiation \cite{Balasubramanian:1996xf}; with decay
rates in hand, we can then discuss the reheating process from thermal
strings to SM radiation.  We find that long open strings redshift like
matter at high densities, and the open string decays reheat the
radiation to a temperature of about the SM string scale.  Open strings
out of thermal equilibrium also could play a role in cosmology, and we
make some qualitative comments about their evolution.

Finally, we conclude by discussing some possible signals of the low SM
string scale, from modifications to the inflationary CMB spectrum to
relics of a possible open string phase.  Exploring these potential
signals is an important direction for future research, especially
given the sensitivity of upcoming CMB experiments.

We give a detailed description of our conventions in appendix
\ref{s:conventions}, for the interested reader.

\section{The naive setting for inflation}\label{s:naive}

To start, we will give a brief review of inflationary models in flux
compactifications in a naive form.  That is, we will describe the
vacuum state of the compactification and then add an inflationary
mechanism as a perturbation of this vacuum. Following our review, we
will argue that this sort of naive addition of inflation to vacuum
compactifications cannot be true in scenarios with multiple throats
where there is a significant disparity in the warping of different
throats. Specifically, we will demonstrate that Kaluza-Klein and $\ap$
corrections would necessarily modify the geometry.

\subsection{Vacuum Compactification Geometry}\label{ss:compact}

Here we will give a brief review of the compactifications we will
discuss.  For specificity, we will focus on the best understood
string compactifications, but our basic conclusions should be
rather generic. Therefore, we start by reviewing common features
of warped compactifications, which will be the most important for
us. Throughout this paper, we will refer to the post-inflationary
compactification as the ``vacuum'' or ``ground state'' geometry.
However, the reader should be aware that, in currently understood
models of moduli fixing, these compactifications with a small
positive cosmological constant are only metastable and may
eventually decompactify to 10D or suffer other decay modes (though
with lifetimes much longer than the age of the universe)
\cite{Kachru:2003aw,Frey:2003dm}.

Consider, then, a 10D braneworld compactification in which the
hierarchy between the Planck and weak scales is provided by a warp
factor, as in \cite{Randall:1999ee}. We take the metric to be
\bea
ds^2 &=& e^{2A}\eta_{\mu\nu} dx^\mu dx^\nu+ g^6_{mn}dx^m dx^n
\ ,\
A\equiv A(x^m)\label{warpmetric}\\
&=& e^{2A}\eta_{\mu\nu} dx^\mu dx^\nu+ e^{-2A} \hat{g}^6_{mn}dx^m
dx^n\ .\nonumber\eea
Here, the internal metric is that of a warped Calabi-Yau manifold
--- that is $\hat{g}^6$, appearing in the second line, is
Calabi-Yau \cite{Giddings:2001yu}. The warp factor provides a
hierarchy of scales \cite{Randall:1999ee}; at the SM brane(s),
$e^{A} \approx \msm/M_P$ implies that 4D SM observers have a
fundamental particle physics scale of $\msm$ (we use conventions
in which $A\sim 0$ away from special points of the
compactification).\footnote{See appendix \ref{s:conventions} for
our conventions for the Planck scale, string scale, etcetera.} For
warping to give the full hierarchy, $\msm\sim 1\,
\textnormal{TeV}$.  For an intermediate scale hierarchy (with the
rest provided, for example, by supersymmetry), simply choose
$\msm$ larger than the weak scale. As will become apparent, the
precise value of $\msm$ will not be important for the following
discussion. In a 10D picture, this means that the SM brane is
located at the bottom of a throat in the compactification (see for
example \cite{Giddings:2001yu} for a string construction).
Following the Kaluza-Klein zero-mode for the graviton, the minimal
prescription for cosmology is to replace $\eta_{\mu\nu}$ by a
cosmological metric $g^4_{\mu\nu}$.

Let us examine the hierarchy provided by the warp factor in more
detail.  Since the warp factor rescales the time coordinate, any
localized mode has its mass rescaled by the warp factor, also, as
was originally emphasized for Higgs fields in
\cite{Randall:1999ee}. Direct calculations bear out this
expectation in a number of regimes; for example, it is well-known
that Kaluza-Klein gravitons localized in the SM throat have mass
given by the 10D curvature scale times the warp factor, $m\approx
e^{\asm}/R$ \cite{Randall:1999vf}, where the throat has anti-de
Sitter (AdS) geometry and $R$ is the AdS radius. This relation
holds even when compactifications have large unwarped regions
\cite{Lykken:2000wz}.  Additionally, the tension of long,
semi-classical strings (extended in the external directions) is
rescaled to $\mu=e^{2\asm}/2\pi\ap$, as we can see directly.
Taking a static gauge in which the worldsheet spatial coordinate
is the target space coordinate length,
\be\label{tension4d} S= -\frac{1}{2\pi\ap}\int d^2\sigma
\sqrt{-\det g_{ab}} =-\frac{1}{2\pi\ap}\int d^2\sigma
e^{2A}\sqrt{1+(\del X^\mu)^2+e^{-2A}(\del X^m)^2+\cdots}\ . \ee
Here $(\del X^\mu)^2$ and $(\del X^m)^2$ schematically represent the
fluctuations in the noncompact and compact directions,
respectively. Further $\cdots$ represent additional mass terms due to
curvature, warping, and possibly other background fields.  Hence, we
see that strings oscillating purely in 4D have a rescaled
tension. Interpolating between 
KK gravitons and semiclassical strings, we expect that the throat
produces a (4D) sector of strings for which all the masses are
rescaled by the warp factor.  In fact, a perturbative quantization of
the string has been done in a slightly different warped background in
\cite{PandoZayas:2003jr}, giving the same result. This fact should
lead to similar phenomenological considerations as
\cite{Cullen:2000ef}.  For notational clarity, we define the effective
string length at the SM brane (located at the tip or bottom of the SM
throat, the region of smallest warp factor),
$\lst=e^{-\asm}\sqrt{\ap}\sim 1/\msm$, while $\ap$ will always denote
the 10D value.

To be concrete, we will focus on compactifications of the type
described by \cite{Dasgupta:1999ss,Greene:2000gh,Giddings:2001yu} (see
\cite{Frey:2003tf} for a review and more references), although our
comments should apply qualitatively to any warped string
compactification.  An important feature of these compactifications is
that supergravity 3-form flux generates a potential for many of the
compactification moduli, fixing their expectation values.  The
remaining moduli, including the size modulus $u$,\footnote{Again, see
definitions in appendix \ref{s:conventions}.\label{f:append}}
can be stabilized by
nonperturbative superpotentials generated by D-brane instantons or
gaugino condensation \cite{Kachru:2003aw,Denef:2004dm} or potentially
by $\ap$ corrections to the supergravity
\cite{Becker:2002nn,Balasubramanian:2004uy}.\footnote{These
nonperturbative and $\ap$ corrections will modify the geometry we
describe here, but those effects are subleading.  The recent work of
\cite{Giddings:2005ff} provides the first step toward incorporating
these corrections.} Since the VEVs of the stabilized moduli are
model-dependent, we can imagine tuning their values to determine what
regions of parameter space produce specific physics, and we will
typically leave the moduli unspecified.  However, we will sometimes
give numerical estimates, in which case we will assume that the string
coupling $g_s\sim 1/10$ and the volume $e^{4u}/g_s\sim 1-10^3$, which
includes the example of \cite{Kachru:2003aw} --- the combination
$e^{4u}/g_s$ actually corresponds to the real part of the K\"ahler
modulus in the 4D EFT \cite{Giddings:2001yu}.

Recall from (\ref{warpmetric}) that the internal metric is a warped
Calabi-Yau manifold, i.e., $g^6_{mn}=e^{-2A}\hat{g}^6_{mn}$ with
$\hat{g}^6$ Calabi-Yau. The warp factor is negligible, $A\sim 0$, away
from singularities of the Calabi-Yau, and the geometry develops
(locally AdS) throats near the singularities. This warp factor is
sourced by the 3-form fluxes, so the final geometry will depend on the
flux quantum numbers $n_f$ (see footnote \ref{f:append}).
The most studied form of the throats
have roughly the geometry of \cite{Klebanov:2000nc,Klebanov:2000hb}
and are topologically deformed conifolds. The amount of deformation is
controlled by one of the moduli of the Calabi-Yau, and the flux
stabilizes the modulus.  (In an abuse of language, we will continue to
call this scalar a modulus, even though it has a potential, because it
will be lighter than other scales we consider.)  From the current
understanding of the dimensional reduction
\cite{deAlwis:2003sn,Buchel:2003js,deAlwis:2004qh,Giddings:2005ff},
the flux-induced mass of the modulus should be $m\sim g_s n_f/\lst
(e^{-3u}+\cdots)$, where $n_f$ is the number of flux quanta in the
throat (defined more precisely in appendix \ref{s:conventions}) and
$\cdots$ represent possible corrections due to derivatives of the warp
factor.  Most likely, the modulus mass should be between this scale
and the KK mass scale, $g_s n_f e^{-3u}/\lst\lesssim m\lesssim
e^{-\asm}/R$, simply because the derivatives of the warp factor set
the scale $R$.  Other throat geometries (based on other possible
singularities) have recently been studied in
\cite{Cascales:2003wn,Franco:2004jz,Herzog:2004tr,Franco:2005fd}.  In
many cases, the region of small warp factor ($A\ll 0$) has been argued
to be similar to the deformed conifold. In particular, in those cases,
there are deformation moduli which are also stabilized by supergravity
flux. Although there may be other, more exotic throat geometries
possible (e.g., a geometric resolution seems to be lacking for
cascades ending with dynamical supersymmetry breaking
\cite{Berenstein:2005xa,Franco:2005zu,Bertolini:2005di}), our
discussion will apply qualitatively for a large class of them.

There are a number of possibilities for the SM branes.  The
simplest (though not realistic) choice is a stack of D3 or
$\overline{\textnormal{D}3}$-branes.  The antibrane positions are
stabilized at the bottom of the SM throat by the warp factor
\cite{Kachru:2002gs,Kachru:2003aw}, and D3-brane positions can be
stabilized by nonperturbative physics \cite{Kachru:2003sx}.
Additionally, if there is an orbifold fixed point at the center of
the throat, the SM branes can be pinned to the fixed point.  More
realistic gauge theories can arise in that way
\cite{Cascales:2003wn}. Another alternative is to consider
D3/$\overline{\textnormal{D}3}$-branes at the bottom of the throat
intersecting with D7-branes extending through the throats. Much is
known about building realistic models from brane intersections;
for example, see
\cite{Uranga:2003pz,Lust:2004ks,Blumenhagen:2004vz} for reviews.
In that case, the SM modes are the D3-brane open strings as well
as open strings stretching between the D3-branes and D7-branes.
The key point for us is that the SM modes will all be localized on
branes at the bottom of the SM throat.

Once again, let us emphasize that, although we are focusing on a
particular type of compactification, we expect our results to
generalize easily.


\subsection{String Implementations of Inflation}\label{ss:infrev}

We will now review several approaches to inflation within the
context of the compactification models discussed above.  We should
note that all of these inflationary mechanisms are typically
treated as small perturbations of the compactification geometry in
the literature.  In this paper, we will mostly discuss the
approach known as brane inflation, but we will also mention two
other inflationary mechanisms very briefly.

The most popular approach to embedding inflation in string theory is
brane inflation
\cite{Burgess:2001fx,Burgess:2001vr,Garcia-Bellido:2001ky,Jones:2002cv,
Burgess:2003qv}.  In this type of model, the inflationary potential is
provided by the Coulomb attraction between spacetime-filling branes
and antibranes.  For brane inflation to avoid destabilizing the
compactification, the moduli must be stabilized at a sufficiently high
mass scale (and in a deep enough potential well); \cite{Kachru:2003sx}
showed that the compactifications described in the previous section
can provide a suitable framework for brane inflation.\footnote{For
recent refinements of this discussion, see \cite{Giddings:2005ff}.} In
addition, the requisite $\overline{\textnormal{D}3}$-branes are fixed
at the bottom of a throat by the warp factor, and the length of the
throat allows the interbrane potential to be flat enough to support an
accelerating cosmological expansion \cite{Kachru:2003sx}.  One
difficulty of brane inflation scenarios as described by
\cite{Kachru:2003sx} is a supergravity $\eta$ problem; in the most
basic setting for brane inflation, it seems that some fine-tuning is
necessary to get a small enough slow-roll parameter $\eta$.  This
$\eta$ problem has been the focus of much recent work (see, for
example,
\cite{Firouzjahi:2003zy,Burgess:2004kv,Berg:2004ek,Chen:2004gc,
Berg:2004sj}).\footnote{Within slow roll inflation it is possible to realize
assisted inflation~\cite{Liddle:1998jc,Copeland:1999cs,Jokinen:2004bp}
with the help of multiple branes~\cite{Mazumdar:2001mm,Cline:2005ty},
or with multiple membranes in strongly coupled heterotic
M-theory~\cite{Becker:2005sg}, or with
multiple-axions~\cite{Dimopoulos:2005ac}, which can ameliorate the
supergravity $\eta$ problem to some extent.}  We will, however,
simply assume that $\eta \ll 1$ without worrying about whether that
assumption requires fine-tuning or not.  Our interest will be
elsewhere.

The potential provided by $N$ antibranes works out to be
\cite{Kachru:2002gs}
\be
\label{d3barV}
V=\frac{N}{\pi} g_s^3 e^{-12u} e^{4\ainf} M_P^4\ ,
\ee
where the subscript ``$\mathnormal{inf}$'' indicates for the warp
factor at the bottom of the inflationary throat.  Henceforth, we
will set the number of antibranes to $N=1$ for the following
reason: unless the wandering D3-branes are tied together in some
way, the last stage of inflation will be described by the dynamics
of one D3-brane. Therefore, inflation will end with the
annihilation of a single D3/$\overline{\textnormal{D3}}$ pair,
decreasing $V(N)$ to $V(N-1)$. However, $V(N-1)$ is already
included in the post-inflationary cosmological constant, so we
only consider one of the antibranes as driving inflation.

Relating the potential to the inflationary Hubble scale $V=3M_P^2
H^2$, we find the warp factor to be
\be
\label{warpinflate}
e^{\ainf} = \left(3\pi\right)^{1/4} \left(\frac{e^{4u}}{g_s}\right)^{3/4}
\left(\frac{H}{M_P} \right)^{1/2}\approx
\mathcal{O}(1-10^3)\left(\frac{H}{M_P}\right)^{1/2} \ .
\ee
For values of the Hubble constant
\be\label{hrange}
\frac{H}{M_P}\approx \mathcal{O}(10^{-8}-10^{-5})\,,
\ee
the warp factor covers a wide range from $10^{-4}$ (as given in
\cite{Kachru:2003sx}) to nearly unity. Note that
(\ref{warpinflate}) gives an interesting bound on the
compactification moduli in brane inflation: $e^{4u}/g_s\lesssim
(M_P/H)^{2/3}$ since $\ainf \leq 0$.

In a sense, this warp factor is comparatively high, in that the
effective string mass $1/\lss=e^{\ainf}/\sqrt{\ap}$ along with the
associated Kaluza-Klein scale $e^{\ainf}/R$ are considerably
higher than the inflationary Hubble scale:
\be\label{hubbleell} \frac{1}{\lss} = \left(12\pi^3
g_s\right)^{1/4} \left(\frac{M_P}{H}\right)^{1/2} H\gg H\ .\ee
From the point of view of effective field theory, this high value
for the warp factor is reassuring, since it implies that string
and KK corrections will be suppressed.  Additionally, it makes
sense to think of the antibranes as probes on the compactification
manifold.  On the other hand, if we are interested in signals from
string theory, having $\lss H\ll 1$ means that corrections to the
CMB are highly suppressed and most likely
undetectable.\footnote{In brane-antibrane inflation, it was
suggested that detectable non-Gaussianities may be produced after
the end of inflation due to the tachyonic instability triggered by
the open string modes. Such large non-Gaussianity produced can be
helpful in constraining not only the string scale but also the
string coupling~\cite{Enqvist:2005nc,Huang:2005nd}.}

Brane inflation also has a natural exit; when the branes get within
about a string length of the antibranes, a string mode stretched
between them becomes tachyonic, resulting in brane annihilation and
reheating as in hybrid inflation
\cite{Cline:2002it,Sarangi:2002yt,Sarangi:2003sg}.  We will discuss
reheating from brane inflation in more detail in section
\ref{s:warpreheat}.

A related inflationary scenario is the D3/D7 model discussed in
\cite{Herdeiro:2001zb,Dasgupta:2002ew,Dasgupta:2004dw}.  In these
models, worldvolume flux on D7-branes plays the role of antibranes,
and the D3-branes are attracted to the D7-branes.  In the ending stage
of inflation, the D3-brane becomes bound to the D7-brane as an
instanton of the worldvolume flux, and the potential is also of the
hybrid inflation type.  One advantage of D3/D7 inflation is that a
shift symmetry exists in many cases, which serves to flatten the
inflationary potential \cite{Hsu:2003cy,Hsu:2004hi}.

Another related inflationary mechanism is warped tachyonic inflation,
recently proposed by \cite{Cremades:2005ir}.  In this mechanism, a
non-BPS D-brane (presumably wrapping a cycle of the compactification
manifold) decays, much as the brane and antibrane annihilate in brane
inflation.  Localizing the non-BPS brane in a warped region can
naturally satisfy slow-roll conditions.

The final mechanism we mention is known as racetrack inflation,
which makes use of the nonperturbative potential that stabilizes
some of the compactification moduli \cite{Blanco-Pillado:2004ns}.
The idea is that sufficiently general nonperturbative
superpotentials give rise to saddle points in the effective
potential, where the slow-roll parameters become small.  We should
emphasize that the inflaton is a compactification modulus, such as
$u$, so the internal space changes throughout inflation.  This
scenario could possibly lead to nonstandard cosmology during
inflation.

\subsection{Problems with the Simple Picture}\label{ss:problems}

Let us now return to the SM sector, recalling that we are interested
in models in which warping provides some significant contribution to
the SM hierarchy.  In that case, as we saw above, strings at the tip
of the SM throat have an effective mass scale of $1/\lst \sim \msm$
(at whatever intermediate scale we choose).  Similarly, the
Kaluza-Klein mass scale is near $\msm$.  However, during inflation, we
know that SM scale can be much lower than the Hubble scale, $H\sim
10^{13}\,$GeV!  Even so, the usual approach to inflation in these
compactifications is through 4D effective field theory and naively
assumes that the only modification of the compactification geometry is
through the replacement $\eta_{\mu\nu}\to g^4_{\mu\nu}$ of the 4D
Minkowski metric with an FRW metric.  We can quickly see how such a
large Hubble scale leads to inconsistencies in this naive model.

Even at the level of the classical equations of motion, this naive
approach already runs into problems.  If we consider a 5D
Randall-Sundrum model as a proxy for a full 10D compactification,
\cite{Lesgourgues:2000tj,Kanno:2004yb} have shown that the naive
replacement $\eta_{\mu\nu}\to g^4_{\mu\nu}$ breaks down precisely when
$H\gtrsim \msm$.  Heuristically speaking, we would expect just such an
effect because the KK gravitino masses are $\sim \msm$, so higher
dimensional gravity should become important as $H\gtrsim \msm$.  In
particular, the time and space coordinates will have different warping
in the extra dimensions, so the 4D Hubble rate will vary over the
compactification, which can have interesting cosmological consequences
\cite{Mazumdar:2000gj,Mazumdar:2000ch}.  The asymmetric warping for
the time and space coordinates also leads to a significant violation
of Lorentz invariance in the effective field theory.  We would expect
large corrections to the CMB in that case.

However, there are other, even more drastic corrections to the 10D
geometry from string physics.  The most straightforward way to see how
the SM strings lead to an inconsistency is to consider curvatures.  We
can already see from a 4D point of view that higher-derivative
corrections to gravity will become important because $\lst H\gg
1$. We, should, however, look at the 10D curvatures, since we are
interested in the compactification geometry.  Without evaluating the
entire $R^4$ correction to IIB supergravity, we note that the Ricci
scalar is already large.  For a metric of the form (\ref{warpmetric}),
the 10D curvature is
\be\label{ricci} R = 12H^2 e^{-2A} +R_6
-8\Del_6^2 A -4(\Del_6 A)^2 = 12H^2 e^{-2A} +e^{2A}\left( 2\hat\Del^2
A+8(\hat\Del A)^2\right)\ ,
\ee
where $\Del_6$ ($\hat\Del_6$) denotes the covariant derivative for
the 6D metric $g^6_{mn}$ ($\hat{g}^6_{mn}$). In the second
equality, we have used the fact that $\hat{g}^6_{mn}$ is
Calabi-Yau (and therefore Ricci-flat).  Near the bottom of the SM
throat, the curvature is dominated by $H^2e^{-2A}$ and the
derivatives of the warp factor are small in comparison. From
(\ref{warpinflate}), we have then
\be \label{curve} R\simeq 12H^2e^{-2\asm} =\frac{M_P}{3\pi}
\left(\frac{e^{4u}}{g_s}\right)^{-3} e^{4\ainf-2\asm}\ . \ee
Hence if the SM throat is too strongly warped, the curvature scale
will be well above the Planck scale as well as the (10D) fundamental
string scale. In this case, the entire tower of higher derivative
corrections should be important.  In fact, the tip region of the SM
throat would need some sort of nonperturbative string description ---
a worldsheet CFT probably would not be sufficient because the
curvature is beyond the 10D Planck scale as well as the string scale.
Such high curvatures would also lead to a breakdown of the effective
field theory from the 4D point of view.

There is an additional, more subtle way to see that the 4D effective
field theory will break down, because, during inflation, the Hubble
scale exceeds the effective tension of strings localized at the tip of
the SM throat. In particular, any comoving observer will see a
prolific creation of strings out of the vacuum during inflation, which
is similar to the phenomenon discussed in
\cite{Gubser:2003vk,Gubser:2003hm,Friess:2004zk}.
Heuristically, the
calculation is as follows: The Wick rotation of de Sitter spacetime in
global coordinates is an $S^4$ of radius $1/H$.  In this geometry,
there is a worldsheet instanton given by a (Euclidean) string wrapping
an equatorial $S^2$. The action is just given by product of the area
of the $S^2$, $4\pi/H^2$, and the (effective) tension: $S_E =
2/H^2\lst^2=2/\ap\,(H^2e^{-2\asm})^{-1}$, which from the discussion
around (\ref{curve}) is small in the present situation.  Now, the
instanton can be Wick rotated back to Minkowski signature by splitting
the $S^2$ along an equator. With this analytic continuation, we see
that the instanton creates a worldsheet filling a $dS_2$ subspace of
the original universe! Thus, we can interpret the instanton as
creating long strings. The nucleation rate per unit volume of these
strings is $e^{-S_E}$ (times factors usually of order unity), which is
not suppressed. Hence we find the advertised prolific production of
strings. In another language, the comoving observer should see a
thermal bath at the Gibbons-Hawking temperature $H/2\pi$. However,
this temperature is well above the effective Hagedorn temperature
$T\sim 1/\lst$ of the strings localized in the SM throat. Hence the
thermal bath should be composed largely of long strings (actually, it
is likely that there will be many black holes, as well, at such a high
temperature).

Of course, all of these difficulties will also arise due to effective
strings formed by D-branes wrapping cycles in the compactification if
those D-branes are located at the tip of the SM throat.  For instance,
D1-branes at the tip of the SM throat will also have a low string
scale compared to Hubble.  For simplicity, though, we will focus on
fundamental string physics in this paper.


\section{Almost String-Scale Inflation}\label{s:stringinflate}

Above, we have found that a naive approach of string inflation leads
to inconsistencies for scenarios where the warping differs
significantly between the inflationary and SM (or any additional)
throats. However, in this section, we will argue that a more
sophisticated treatment leads to a picture where the highly warped SM
throat becomes essentially ``self-repairing'' during inflation.  We
will find a self-consistent picture involving 4D effective field
theory and a modification to the compactification geometry.  First, we
will examinine the consequences of 4D effective field theory during
inflation and interpreting them in terms of 10D physics.  Then we will
see that the emergent 10D picture is still consistent with 4D
effective field theory with controlled $\ap$ and higher dimensional
corrections.  At the same time, we will see how, at least in the
correct region of parameter space, those corrections could become
observable.

We might note that
\cite{Turok:1988pg,Sanchez:1990cw,Gasperini:1991xg,Gasperini:1991rv}
have considered inflationary scenarios with the Hubble scale at or
above the string scale, giving rise to an ``unstable'' phase of
strings. Our ``self-repairing'' scenario is clearly distinct in that
we do not encounter any such exotic stringy phase (during inflation).

\subsection{Hubble-Induced Masses}\label{ss:hinducedm}

To begin, we assume the validity of effective field theory during
inflation, and we will see that it is generically impossible for
any scalar to have a mass parametrically lower than the Hubble
scale.  In the following section, we will then argue that this
fact implies that during inflation the SM throat geometry is
modified from its true vacuum.

One argument to this effect has been known for some time in
supersymmetric models for cosmology
\cite{Dine:1983ys,Bertolami:1987xb,Copeland:1994vg,
Dine:1995uk,Dvali:1995mj,Dine:1995kz} (see also
\cite{Enqvist:2003gh}), which we review here.  Consider some
supergravity theory with a light scalar $\phi$ (which by abuse of
language we will call a modulus), although our argument holds true
even for an inflaton.  The basic point is that the $\N=1$ supergravity
potential
\be\label{sugrapot}
V \propto e^{\mathcal{K}} \left(
\mathcal{K}^{i\b\jmath}D_i W D_{\b\jmath} \b W-3|W|^2\right) +|D|^2\,,
\ee
contains many cross-couplings between any such $\phi$ and the
inflaton energy density.  For example, assume that $\phi$ has a
minimal K\"ahler potential $|\phi|^2/M_P^2$ and that the inflaton
potential comes from the F terms.  Then the total potential
contains a Hubble induced mass term of roughly
\be
\label{Hmass}
\frac{1}{M_P^2}|\phi|^2 V_{\mathnormal{inf}} \approx C H^2 |\phi|^2\ .
\ee
In fact, \cite{Dine:1995uk,Dine:1995kz} give many possible such Hubble
contributions to the effective potential of $\phi$, leading to an
effective mass term as above where $|C|= O(1)$ but $C$ may have either
sign. If $\phi$ were the inflaton, then it would suffer the the
supergravity $\eta$ problem during inflation, where $|\eta|\equiv
M_p^2V^{\prime\prime}/V\sim \mathcal{O}(1)$. As it turns out,
\cite{Kachru:2003sx} showed that this type of $\eta$ problem can
generalize even to $D$-term inflation, in particular to the case of
brane inflation.  In this paper we assume that the $\eta$ problem for
the inflaton has been tackled by fine tuning or some other
inflaton-specific mechanism. Fine tuning could also render the mass of
the modulus $\phi$ to be light compared to the Hubble mass ($C\ll
1$).\footnote{For a no-scale type K\"ahler potential $C=0$. At one
loop, corrections could be induced, but generically they are small,
$|C|\ll 1$~\cite{Gaillard:1995az}.}  However, we will see below that
the percent level fine tuning that is sufficient for the inflaton
$\eta$ problem does not change our story significantly when applied to
the modulus $\phi$.

Additionally, we might be interested in a negative Hubble induced
mass, $C<0$, when it is possible to stabilize the modulus at a finite
VEV \cite{Dine:1995uk,Dine:1995kz}.  Besides the soft SUSY breaking
mass term (\ref{Hmass}), there could be other contributions from the
nonrenormalizable superpotential contribution, i.e., $W\sim
\lambda\phi^{d}/M^{d-3}$ with some cut-off scale $M$.  Such a
contribution would lead to a self interacting potential
\be
|\lambda|^2\frac{\phi^{2d-2}}{M^{2d-6}}\,,
\ee
which can stabilize the modulus to a false vacuum with a finite energy
density, where $\lambda\sim \mathcal{O}(1)$ and $d\geq 3$.  During
inflation the VEV of the modulus will be given by
\be\label{infvev}
|\phi|\simeq\left(\frac{C}{(d-1)\lambda}HM^{d-3}\right)^{1/(d-2)}\,.
\ee
For example, such nonrenormalizable potentials would arise from
integrating out the heavy KK modes.  Therefore, the cut-off $M$ would
be given by the Kaluza-Klein or string scale, which is essentially
determined by the behavior of $\phi$.  In the end, we expect $M\sim H$
during inflation, so the VEV is bounded by $H$.  Also, the mass at the
new VEV will also be of Hubble scale.

Next we consider a complementary argument that moduli will generically
develop Hubble scale masses during inflation. If our light scalar has
a mass $m \lesssim H$ in its effective potential, its quantum
fluctuations during inflation develop a (steady-state) VEV. Strictly
speaking, for a massless scalar in a de Sitter background, the two
point correlation function grows indefinitely for long wavelength
fluctuations \cite{Vilenkin:1983xp}
\be
\langle \phi^2\rangle \approx \frac{1}{2\pi^3}\int_{H}^{He^{Ht}}
d^3k|\phi_{k}|^2\approx \frac{H^3}{4\pi^2}t\,.
\ee
This result can also be obtained by considering the Brownian motion of
the scalar field. For a massive field, one does not obtain an
indefinite growth of the variance of the long wavelength fluctuations,
but \cite{Vilenkin:1982wt,Vilenkin:1983xp,Linde:1982uu}
\be\label{vev1}
\langle \phi^2\rangle = \frac{3H^4}{8\pi^2m^2}
\left(1-e^{-(2m^2/3H)t}\right)\longrightarrow \frac{3H^4}{8\pi^2m^2}
\ee
after sufficient e-foldings.  In the limiting case when $m\rightarrow
H$, the variance goes as $\langle \phi^2\rangle \approx H^2$.

Let us now suppose that $\phi$ has self-interactions, including
generically a quartic term $\lambda\phi^4$.  Expanding the action for
particle-like fluctuations $\delta\phi$, we find an effective mass
term
\be\label{effmass} \frac{1}{2} m_{\mathnormal{eff}}^2 \delta\phi^2
= \frac{1}{2}m^2\delta\phi^2 +
6\lambda\langle\phi^2\rangle\delta\phi^2 \ee
From (\ref{vev1}), the contribution from the interaction will in fact
dominate the bare mass unless $\lambda<m^4/H^4$, which is unexpectedly
small in our case.  Solving (\ref{vev1}) for the effective mass, we
obtain
\be\label{vev2} m^2_{\mathnormal{eff}} \approx
\frac{3}{\pi}\sqrt{\frac{\lambda}{2}}\, H^2\ ,\ \ \langle
\phi^2\rangle \approx \frac{H^2}{4\pi\sqrt{2\lambda}}\ . \ee
So we see that a self-interaction generates a Hubble-scale induced
mass for a light scalar during inflation. This argument holds even if
the scalar VEV is shifted by corrections such as
(\ref{infvev}). Further, this argument generalizes to
nonrenormalizable interactions.

We stress that we are \textit{not} taking the view that the light
scalar starts in its true vacuum and then moves away, either due to an
induced classical potential or quantum fluctuations.  Instead, we
imagine that the modulus has some (arbitrary) initial condition during
some inflationary phase and then rolls toward a minimum of the
potential over a few e-foldings.  However, inflationary physics,
including quantum fluctuations, keeps $|\phi|\sim H$ over a given
patch.  The modulus $\phi$ only approaches its true (``bare'') ground
state after the end of inflation.

Finally, we remind the reader of the potential role in cosmology for
such a light scalar.  For instance, its presence during inflation
typically gives rise to isocurvature fluctuations. If the modulus is
sufficiently long lived (longer than the inflaton), then its decay
could lead to the reheating the Universe and the conversion of
isocurvature fluctuations into adiabatic density perturbations
\cite{Enqvist:2002rf,Enqvist:2003mr,Mazumdar:2003bs,Enqvist:2003qc}.
Within the MSSM there are $300$ such moduli which can play an
important role in cosmology \cite{Enqvist:2003gh}.  We will return to
reheating from moduli in warped compactification scenarios in section
\ref{ss:modulus}.

\comment{ The fluctuations of the scalar mean that any light
modulus can become a homogeneous condensate with a nonvanishing
VEV after couple of e-foldings of inflation, homogeneous enough to
be treated as a classical field within one Hubble radius. However,
if the condensate is heavy, $m\gg H$, then the field will roll
down quickly to the bottom of the potential within a Hubble time
and will eventually settle down at the minimum after few
oscillations. Such a modulus is unlikely to play any significant
role in cosmology.

On the other hand, if $m\lesssim H$, as we have, then the condensate
will store a finite energy density for disposal. The same is true for
a negative Hubble induced mass to the moduli. In both the cases the
maximum energy density stored in the modulus is of order
$\rho_{\phi}\lesssim\mathcal{ O}(1) H^4$.  Clearly, this is less than
the inflationary energy density, so it does not affect inflation.  The
main point, though, is not the value of the energy density but that
this modulus can contribute to reheating of the SM branes.  As we
shall see, a similar conclusion can be reached while modeling the SM
throat as a black brane in an alternate proposal.}


\subsection{Ten-dimensional Interpretation}\label{ss:10dinterp}

Recall from section \ref{ss:compact} that the vacuum state geometry of
the SM throat suppresses the mass of localized degrees of freedom by a
factor of $e^{\asm}\sim \msm/M_P$.  Therefore, since the maximum
fundamental mass scale is $M_P$, the greatest mass the deformation
modulus should take is $\msm$; this is just the Randall-Sundrum
hierarchy.  However, by the arguments of section \ref{ss:hinducedm}
above, this modulus should have a mass of approximately $m\sim H$
during inflation if the 4D effective theory is valid.  Notice the
tension in the two statements. Either effective field theory must
break down, or the SM throat geometry must be modified in some manner
consistent with effective field theory.  We assume that inflation does
indeed induce a Hubble scale mass for the SM throat deformation
modulus, precisely as in effective field theory.  What we now argue is
that the use of effective field theory is consistent from the 10D
point of view and that all the problems of section \ref{ss:problems}
are resolved.\footnote{We discuss some alternative possibilities in
appendix \ref{s:alternatives}.}

The effective theory itself suggests what happens from the 10D point
of view during inflation.  In the effective theory, all scalars
fluctuate during inflation, which was part of our argument for
Hubble-scale effective masses.  Fluctuations of the SM throat
deformation modulus actually cause the throat geometry itself to
fluctuate, so we should not expect the SM throat to remain in the
vacuum configuration.  Additionally, the Kaluza-Klein gravitons, which
are also light in the vacuum compactification, should fluctuate as
well, further modifying the throat geometry.  Unfortunately, we do not
have a way to describe the 10D geometry in complete detail.  What
seems clear, though, is that the warp factor at the tip of the SM
throat must be lifted to at least $e^{\asm}\sim H/M_P$.  This lifting
will allow the deformation modulus to acquire a Hubble scale mass.
The reader should note that this value is still much smaller than the
inflationary throat warp factor $e^{\ainf}$ (\ref{warpinflate}), which
we assume remains essentially unchanged during inflation.

Hence we assume that the warp factor lifts up, so the effective mass
of the deformation modulus in the SM throat blueshifts to $m^2=CH^2$,
where $C\approx 1$. Since all of the masses relevant for this throat
are proportional to the warp factor, the effective string and
Kaluza-Klein scales are also raised. To determine the warp factor more
precisely, we may further assume that the modulus mass is given by
usual formula for a flux-induced mass. Following the discussion in
appendix \ref{s:conventions}, we then have
\be\label{newwarp} e^{\asm}\approx\frac{\sqrt{C}}{g_s n_f}
e^{3u}\sqrt{\ap}H=\frac{\sqrt{C}}{g_s^2n_f} e^{6u}\frac{H}{M_P}\
.\ee
Note that the second equality uses the fact that change in the
geometry of the SM throat will leave the internal volume
(\ref{warpedV}) essentially unchanged. Hence, with reasonable values
of the parameters (e.g., $g_s\sim 1/10$, $n_f\sim 10$, $e^{4u}/g_s\sim
10^3$ as given in \cite{Kachru:2003aw}), the warp factor is somewhat
larger than the naive estimate above, i.e., $e^{\asm}\approx 10^{4}
H/M_P$. With (\ref{newwarp}), we can also estimate the effective
string scale in the SM throat during inflation to be\footnote{In
general, the SM throat geometry will vary with time, and so in the
following, we will use $\lst$ to denote the instantaneous effective
string scale at any given stage of the throat's evolution. However, we
will reserve $\msm$ to denote the effective string mass in the SM
throat only when the latter has reached its ``vacuum" configuration.}
\be\label{inflatestring}  \frac{1}{\lst^2} \approx \frac{C
e^{6u}}{n_f^2 g_s^2} H^2\ \ \ \textnormal{or}\ \ \ \lst H \approx
\frac{n_f g_s}{\sqrt{C}} e^{-3u}\lesssim 1\ . \ee
In fact, with the parameter values above, we find $\lst H\approx
10^{-1.5}$. Furthermore, if the compactification scale is of the
order of the 10D string scale (i.e., $e^u\rightarrow1$), we find
that $\lst H$ approaches unity.\footnote{Also, if derivatives of
the warp factor contribute appreciably to the modulus mass, $\lst
H$ might be somewhat closer to unity than given in
(\ref{inflatestring}).} Similarly the mass scale for KK modes
localized in the SM throat is given by
\be\label{inflatekk} m_{KK}^2\simeq \frac{e^{2\asm}}{R^2} \approx
\frac{C e^{6u}}{n_f^2 n_R g_s^3} H^2 \ee
where the AdS curvature at the tip is roughly $R^2\sim g_s
n_R\ap\gtrsim \ap$ in string units --- see appendix
\ref{s:conventions}. As usual, the latter inequality ensures that
the KK mass scale is slightly smaller than the effective string
scale.

Perhaps the most uncertain assumption above was that the mass of the
deformation modulus had the scalings of a typical flux-induced mass. A
slightly more conservative estimate would be that the modulus mass
corresponds to that of the KK modes localized in the throat. In this
case, the hierarchy between the effective string and Hubble scales is:
$\lst H \approx \sqrt{\alpha'}/\sqrt{C}R$.  Hence, the separation of
scales is independent of the compactification volume and so may be
smaller than estimated above. We must require $\sqrt{\alpha'}/R<1$ for
the SM throat in its ground state in order that supergravity
calculations are reliable. As the RR flux determining R is fixed by a
quantization condition, one might expect this inequality is maintained
in the fore-shortened throat produced during inflation.

In any event, either approach seems to yield $\lst H\lesssim 1$.
Hence it would seem that the compactification is just within the realm
of validity for 4D effective field theory. Higher dimensional and high
curvature corrections are somewhat suppressed (and, in fact, the
suppression is by discretely tunable parameters), so we have a
self-consistent picture for the SM throat during inflation.  That is,
the SM throat has some initial condition which is \textit{not} its
true ground state, and, after a few e-foldings, the throat settles
into an inflationary state with $e^{\asm}\sim H/M_P$. In fact, this
evolution simply mimics the behavior expected for a scalar field
during inflation from the point of view of 4D effective field theory.

Additionally, in the correct range of parameter space, $\lst H$ is not
so much less than $1$ so that corrections to the CMB may be
detectable.  In fact, with the parameter values considered above, we
expect that higher dimensional and string theoretic physics may
correct the 4D effective theory at about a percent level or even
somewhat higher. In terms of the CMB, the corrections may appear as a
nongaussianity (possibly) or, even more likely, as a deviation from
scale-invariance.  However, we leave determining the precise form of
the corrections for the future.  One direction that may be fruitful to
pursue is determining the geometry of the compactification throat in a
simplified model; as indicated in
\cite{Lesgourgues:2000tj,Kanno:2004yb}, the time and noncompact space
directions will have warp factors that differ by $\mathcal{O}(\lst H)$
corrections, and perhaps a modified cosmology can be extracted from
the accompanying violation of Lorentz invariance.

\section{Reheating of Standard Model Throat}\label{s:warpreheat}

During inflation, the inflaton dominates the energy density of the
universe, and any radiation or matter density redshifts to
negligible values very quickly.  How the energy density ends up in
SM degrees of freedom after inflation is the question of
reheating, and it seems a particularly challenging problem in
brane inflation models, where the inflationary and SM branes can
be separated both geometrically and by large potential barriers
(due to the warp factor). Fortunately, the inflationary and SM
throats can communicate
\cite{Dimopoulos:2001ui,Dimopoulos:2001qd}, and the SM can reheat
\cite{Barnaby:2004gg}.  In this section, we will review the
reheating mechanism of \cite{Barnaby:2004gg} and give a more
detailed calculation of the reheating rate and final reheat
temperature of the SM throat.\footnote{See also the recent
discussion in \cite{Kofman:2005yz,Chialva:2005zy}.} 
Then we discuss the dynamics
of the SM throat itself during the era following inflation and
comment on what effects those dynamics can have on reheating.
Based on our reheating calculations, we will find that strings
will likely have an important role in reheating the light SM
fields.  The key point is that, in a wide range of parameter
space, the SM sector reheating temperature (as calculated in EFT)
is higher than the Hagedorn temperature of the SM strings. We turn
to the cosmology of those strings in the next section.


\subsection{Reheating from Inflationary Throat}\label{ss:tunnel1}

Reheating begins in the inflationary throat with tachyon condensation
when the brane/antibrane pairs are within a string length of each
other
\cite{Burgess:2001fx,Mazumdar:2001mm,Burgess:2001vr,Cline:2002it}.
Because the tachyon couples to the brane gauge fields, its condensate
actually breaks the gauge groups on the branes as in the Higgs
mechanism.  The rolling of the tachyon also excites the gauge fields
on any remaining branes along with massive closed strings. One of the
attractive features is, of course, the formation of cosmic F- and
D-strings by the Kibble mechanism, as widely discussed in the
literature \cite{Jones:2002cv,Barnaby:2004dz}; their stability (and
its model dependence) has been discussed in
\cite{Copeland:2003bj,Jackson:2004zg}. In any case, we expect that
much of the inflationary energy density will end up as gravitational
modes on a short time scale.  Here, we will review the argument given
by \cite{Barnaby:2004gg}.  Reheating proceeds in a multi-stage
process, and we will examine each step.  Finally, we will give an
estimate of the reheating temperature and energy density.  In the
following, we will use $H_\star$ to denote the Hubble parameter at the
end of inflation.\footnote{We discuss possible generalizations of this
mechanism and reheating in other string embeddings of inflation in
appendix \ref{s:otherreheat}.}

We would also like to note that reheating itself can lead to
constraints on brane inflation models, although our focus is on the
final reheat temperature.  The papers 
\cite{Barnaby:2004gg,Kofman:2005yz,Chialva:2005zy}
have considered graviton production during reheating, and
\cite{Kofman:2005yz} has additionally addressed the production of
other dangerous relics, such as long lived KK modes in the
inflationary throat.

\subsubsection{Brane Annihilation}\label{sss:annihilate}

As we noted in section \ref{ss:infrev}, brane inflation is a stringy
realization of hybrid inflation.  When the brane and antibrane come
within a string length, an open string mode becomes tachyonic, and the
slow roll conditions fail.  Condensation of the tachyon annihilates
the branes, which creates massive closed strings. Although a
homogeneous calculation for D3-branes does not lead to efficient
reheating, tachyon modes with nonzero momentum can lead to more
efficient decay channels \cite{Lambert:2003zr}.  In that case, the
closed strings have typical mass $m\sim 1/\lss g_{s}$ and
nonrelativistic transverse momentum $p_T\sim 1/\lss \sqrt{g_{s}}$,
where again the subscript ``$\mathnormal{inf}$'' indicates the value
at the bottom of the inflationary throat. This intrinsically stringy
annihilation has a time scale of roughly $\lss$, so the total energy
density stored in the closed string sector is of the order of
$\tau\sim 1/\lss^4 g_s$, the brane tension.\footnote{See also
\cite{Sen:2004nf} for a review and further references.}

We also note that the tachyonic instability can also excite nonlinear
gravitational fluctuations, giving rise to significant nongaussianity
in the CMB, which is constrained by present data from WMAP.  In some
situations, these constraints put limits on the inflationary throat
string scale \cite{Enqvist:2005nc}, although the cases we consider are
probably safe because $\lss H_\star\ll 1$.


\subsubsection{Tunnelling Rate}\label{sss:tunnel}

To calculate the rate of decay from massive closed strings to KK modes
with significant wave functions in the SM throat, it is helpful to
think of this decay as a two step process.  First, we note that the
massive closed strings, being nonrelativistic in the compact
directions, will be localized near the tip of the inflationary throat
at the time they decay.  Therefore, this decay should occur at the
local string time scale $\lss$ and should result in KK gravitons
\textit{localized in the inflationary throat}.  The next step is then
the transfer of these gravitons to the SM throat by tunnelling, as has
been discussed in \cite{Barnaby:2004gg} --- see related discussion in
\cite{Kofman:2005yz,Chialva:2005zy}.

Actually, this tunnelling rate has been considered earlier in
\cite{Dimopoulos:2001ui,Dimopoulos:2001qd}, which discuss the
tunnelling of localized gravitons from one AdS throat to another.
Those papers give a tunnelling rate
\be\label{wtunnel} \Gamma_{t} =
\frac{\pi^2 n^4}{16} e^{5\ainf}\frac{1}{R} \ ,
\ee
where $R$ is the curvature scale of the initial throat (and by
assumption, of the final throat) and $n$ is the KK mode number of
the graviton (measured in the initial throat).  $R$ is actually an
effective anti-de Sitter radius, which is $R\sim \sqrt{g_s
n_R\ap}$ (see appendix \ref{s:conventions} for the definition of
$n_R$) and $\ainf$ is given by (\ref{warpinflate}). In terms of
the inflationary Hubble scale $H_\star$, the rate is
\be\label{wdecay1}
\Gamma_{t} \approx 3\pi^2\left(\frac{n}{2}\right)^4
\left(\frac{12\pi^7}{g_s n_R^2}\right)^{1/4}
\left(\frac{e^{4u}}{g_s}\right)^3\left(\frac{H_\star}{M_P}\right)^{3/2}
H_\star\, .
\ee
Taking $n_R\sim 10- 10^2$, $n=1$, and our usual values
$(e^{4u}/g_s)\sim 1 - 10^3$ and $g_s\sim 1/10$, we find that
$\Gamma_{t}$ ranges over $\mathcal{O}
(10-10^{10})(H_\star/M_P)^{3/2}H_\star$, largely due to the strong
dependence on $e^{4u}/g_s$.

Because the rate (\ref{wdecay1}) is very sensitive to the KK mode
number, we should also check that $n\sim 1$ is a reasonable
assumption.  We know that the compact momentum of each KK graviton is
$p_T = f/\lss\sqrt{g_s}$ with $f\sim 1/2$ just from the transverse
momentum of the massive closed strings.  However, the mass (and
transverse momentum) are quantized approximately in units of
$1/\lss\sqrt{g_s n_R}$.  Therefore, the typical KK mode number is
$n=f\sqrt{n_R}$.  For $n_R\sim 10-100$, we have at most $n\sim
5$. Additionally, the KK modes can interact with each other and decay
into lighter KK states with a time scale of order $\lss$.  This decay
competes with tunnelling, so we expect some reduction in $n$ for the
typical tunnelling rate.  Therefore, it seems reasonable to take a
small typical mode number $n\lesssim 3$.

There is another effect, which we have not calculated, which could
possibly suppress the tunnelling rate further. In the 5-dimensional
models used to approximate the KK mode wave functions and the
tunnelling rate between throats, the two distinct throats have the
same AdS curvature scale.  However, in the string compactifications we
study, the different throats have different AdS curvatures.  If the
curvature in the SM throat (at the appropriate value of the warp
factor) is larger than in the inflationary throat radius, then the
tunnelling rate would be suppressed \cite{Langfelder}.  However, we
will assume that the suppression is not great enough to reduce the
rate much below our estimate (\ref{wdecay1}).

For completeness, we can also give the tunnelling rate in terms of
the inflationary throat string scale, which is the form of the
tunnelling rate given in \cite{Barnaby:2004gg}:
\be \label{tunrate}\Gamma_{t} = \pi^2\left(\frac{n}{2}\right)^4
\left(\frac{1}{g_s n_R}\right)^{1/2} e^{4\ainf}\frac{1}{\lss}\ .
\ee
This is heavily suppressed with respect to the string scale, so
the total decay rate for the decay of massive string modes to SM
throat KK modes is determined by this tunnelling rate. Comparing
this rate to the decay rate into massless 4D gravitons imposes
strong restrictions on constructing a viable model --- see archive
version of \cite{Barnaby:2004gg}.

It may be that considering the tunnelling process with a more
sophisticated model reduces this suppression (\ref{tunrate}) and
eases these restrictions. For example, the potential barrier of a
cascading geometry is smaller \cite{Langfelder} than in the simple
AdS model of \cite{Dimopoulos:2001ui,Dimopoulos:2001qd}. Another
possibility comes from the introduction of D7-branes in the SM
construction, as was briefly mentioned at the end of section
\ref{ss:compact}. The purely D7-brane open strings have KK scale
masses and are weakly coupled to the SM fields. However, they may
still provide an interesting channel for tunnelling if the
D7-brane ventures out from the SM throat to the inflationary
throat.


\subsubsection{Excitation of SM Brane Modes}\label{sss:braneexcite}

The ultimate stage of warped reheating is excitation of the SM brane
degrees of freedom, which effectively traps energy density in the SM
sector.  Since the SM branes lie at the bottom of the SM throat, we
expect that the decay of the KK modes into light SM brane modes will
occur over a time $\lst$, the string scale at the tip of the SM
throat.  The key issue, then, is the value of $\lst$ during reheating;
after all, during inflation, $1/\lst\gtrsim H_\star$, while the
geometry should settle to its vacuum state $1/\lst\sim \msm$ long
after inflation.  As we argue in section \ref{sss:evolution}, we
expect that the time dependent string scale is about the Hubble scale
at that time, or $1/\lst(t)\sim H(t)$.  Further, the time scale for
energy to reach the SM throat is the decay rate (\ref{wdecay1}), so we
expect the string scale at reheating to be $1/\lst
|_{\mathnormal{reheat}}\sim \Gamma_t$, unless $\Gamma_t\lesssim \msm$,
in which case $1/\lst\sim\msm$.  In either case, we expect the brane
excitation rate $\Gamma_{\mathnormal{SM}}\gtrsim\Gamma_t$.

However, the above discussion assumes that the KK gravitons interact
perturbatively with the branes.  If there is any coherence in the KK
modes, however, semiclassical effects can be important.  In
particular, there is evidence that bulk modes can reheat the SM brane
through parametric resonance \cite{Diakonos:2004xq,Seahra:2005wk}.
Parametric resonance, or preheating, can be efficient and rapid
\cite{Traschen:1990sw,Shtanov:1994ce,Kofman:1994rk,Kofman:1997yn},
occurring over a time scale set by oscillations of the KK gravitons.
In our case, this time scale is shorter than $\lst$.  Even so, the
thermalization time scale for SM brane modes would be around $\lst$
(at least in effective field theory), so we end up with the same
conclusion as the purely perturbative interaction.


\subsubsection{SM Reheat Temperature}\label{sss:reheattemp}

The final stage of reheating is thermalization of the SM degrees
of freedom. In general, there is a stage with reheating rate
$\Gamma < H(t)$, so the SM modes will redshift appreciably before
thermalizing, and we can consider the reheating to occur when
$\Gamma \sim H(t)$. We have argued that the reheating rate is
determined by the tunnelling rate, so $\Gamma\sim \Gamma_t$.
Assuming that there is instantaneous thermalization of the decay
products, the reheat temperature comes out to be~\footnote{A full
thermalization in field theory case need not be completed during
reheating \cite{Allahverdi:2005fq}, especially in the case of
supersymmetric theories, because $2\leftrightarrow 2$ and
$2\rightarrow 3$ processes, required for
thermalization~\cite{Jaikumar:2002iq}, are mediated via gauge
bosons/gauginos and Higgses. They obtain large masses from the
VEVs of the flat directions of the minimal supersymmetric SM,
which slows down the rate of thermalization. It is plausible to
define a temperature corresponding to the kinetic equilibrium
despite the lack of full thermal equilibrium, in our case we will
treat $T_{rh}$ as a symbolic temperature which would help us
comparing the decay rates from various processes.}
\be\label{rhtemp} T_{rh}\sim 0.1\sqrt{\Gamma M_{P}}\sim
\sqrt{\Gamma_t M_P} \ ,
\ee
which follows by comparing the thermal energy density $\sim T^4$ (for
SM radiation) to the energy driving the expansion $\sim H(t)^2 M_P^2$.

There are two cases we should consider: If $\Gamma_t\gtrsim
\msm^2/M_P$, the reheat temperature is larger than $\msm$ (at the
time of reheating) --- one finds that this is a plausible
situation combining (\ref{inflatestring}) and (\ref{wdecay1}). In
this case, the thermal bath is sufficiently hot enough to excite
fundamental strings (and perhaps D-strings). This means that the
estimate (\ref{rhtemp}) is unreliable, since it depends on field
theoretic relations between the temperature and energy density.
Instead, we should have a truly string theoretic reheating
process.  In that case, thermalization should result in a gas of
long open strings in the Hagedorn regime $\beta\sim\lst$ and
energy density near $\sim H(t)^2M_{P}^2\sim \Gamma_t^2 M_P^2$.
(Again, we are simply assuming that the energy driving the
expansion of the universe is dominated by the open strings.) At
temperatures above $\msm$ we would also excite the KK modes in the
SM throat. However, they need not be in thermal equilibrium with
the SM degrees of freedom because the KK modes do not share the
same charges as the SM. For example, KK modes with high angular
momentum in the compact directions (as in \cite{Kofman:2005yz})
may be localized in the SM throat away from the SM branes, so they
may only be weakly coupled to the SM degrees of freedom. It would,
of course, be interesting to describe the string reheating process
in more detail.  We will discuss the long open string phase and
its cosmological evolution in section \ref{s:decayreheat}.

The second case is that $\msm$ is relatively large and $\Gamma_t$
is small so that $T_{rh}\lesssim \msm$.  In that case, the field
theoretic reheating calculation will be valid, and only SM
radiation (and no open strings or KK modes) will be excited.

\subsection{Reheating from Modulus Decay}\label{ss:modulus}

In section \ref{s:stringinflate}, we described how the deformation
modulus should lift the warp factor in the SM throat during
inflation so that $\lst H_{inf} \lesssim 1$. Our field theoretic
expectations are then that the modulus obtains a VEV of order
$H_{inf}$ and contributes to the energy density with $\rho \sim
H_{inf}^4$. Of course, as the Hubble parameter evolves during the
last few e-foldings of inflation and reheating, so should the
modulus in the SM throat. Here, we will discuss possible scenarios
for the modulus evolution and the consequences for cosmology.

\subsubsection{SM Throat Evolution}\label{sss:evolution}

There are two distinct possibilities, associated with two
possibilities for the Hubble-induced potentials discussed in section
\ref{ss:hinducedm}.  Simply put, the two possibilities correspond to
whether the VEV of the modulus is induced by cross-couplings in the
potential or by quantum fluctuations, as described in section
\ref{s:stringinflate}A.

The case of cross-couplings, in which the potential develops a
tachyonic mass $m^2_{\mathnormal{eff}}\sim -H^2$ near $\phi=0$, is
of importance phenomenologically for the Polonyi problem
\cite{Dvali:1995mj} and baryogenesis
\cite{Dine:1995uk,Dine:1995kz}, and \cite{Dine:1995kz} has already
studied the behavior of a modulus with such an induced potential,
so we will review their conclusions.  Let $\phi_0(H)$ be the
Hubble-induced ground state (\ref{infvev}) for the given value of
the Hubble constant.  Since the potential is induced by the energy
density, it will remain throughout reheating, and $\phi(t)$ will
track $\phi_0(H(t))$ until $H(t)\sim \msm$ for a matter-dominated
epoch. (It is a simple calculation to show that their argument
carries over to the radiation dominated case, as well.) Therefore,
we expect that the modulus $\phi$ should not begin oscillating
around its true vacuum until $H(t)\sim \msm$, which is when the
``bare'' potential should become important --- we expect the bare
or vacuum mass of the modulus to be closer to the KK scale but
there should only be a small separation between these scales.
During the oscillating phase, preheating and perturbative
reheating will proceed as usual, except that the modulus will not
necessarily dominate the energy density.  As a function of time,
this motivates us to expect that $\lst(t) H(t)\sim 1$ for
$H(t)\gtrsim \msm$.

In the second case where quantum fluctuations dominate the VEV,
$\langle\phi\rangle =0$ over many horizons but
$\langle|\phi|^2\rangle\neq 0$.  In fact, because of the two-point
expectation, $\phi$ has a coherent phase and magnitude $|\phi|\sim
H$ over horizon distances $1/H_\star$ at the end of inflation.  In
this case, we expect that $\phi$ will begin to oscillate around
its true vacuum relatively quickly. However, the magnitude of the
oscillation decays as $1/t\propto H(t)$ (disregarding reheating),
and we might expect that the SM throat warp factor only decays
with the magnitude of the oscillation, so $\lst(t) H(t)\sim 1$ for
$H(t)\gtrsim \msm$. Once again, preheating and reheating will
proceed as normal.

Before proceeding, let us note two things.  First, we should take the
preceding discussion of the modulus (and SM throat) evolution with a
grain of salt.  After all, it would not be surprising if there are 10D
or string corrections, and we do not have a complete picture of the
stationary geometry during inflation.  Still, we expect that the
effective field theory gives us a reasonable qualitative picture,
given that the corrections are somewhat suppressed during
inflation. Additionally, when $H(t)\lesssim \msm$, we expect the SM
throat to settle into its true vacuum, so we at least understand the
endpoint of its evolution.  The other point is that we do not expect
the SM throat modulus to make a dominant contribution to the energy
density initially.  The modulus might eventually dominate the energy
density if it decays slowly enough to SM fields (because the energy
density in the modulus redshifts at least as slowly as matter).  This
is reminiscent of the cosmological moduli problem, and we will explore
this point below.


\subsubsection{Reheat Temperature from Modulus}\label{sss:rhtempmod}

Once the modulus oscillates coherently (roughly when the Hubble
parameter is less than the modulus mass $H(t)\lesssim m_\phi$), the
oscillations mimic a pressureless fluid, whose energy density
redshifts as $a^{-3}$, where $a$ is the scale factor.  During this
oscillation, the modulus can decay nonperturbatively. This is due to
the fact that every time the modulus passes through its minimum
$\phi=0$, the velocity of the field changes sign and vacuum
fluctuations are enhanced due to nonadiabatic evolution. This triggers
the fragmentation of the homogeneous condensate, known as preheating
\cite{Traschen:1990sw,Shtanov:1994ce,Kofman:1994rk,Kofman:1997yn}.

Although this initial stage of particle creation could be
explosive, preheating does not lead to a complete thermalization.
Thermalization is achieved via decay of the condensate (assuming
that preheating does not use all the energy of the condensate). If
the effective mass of the decay product is heavier than the mass
of the modulus, $g\phi>m_{\phi}$, where $g$ is some gauge or
Yukawa coupling, the modulus could decay into light SM particles
only through loop diagrams involving heavy particles, with an
effective coupling of the type
$(g^2/\langle\phi\rangle)\phi\bar\psi\gamma^{0}\partial_{0}\psi$,
where $\psi$ is a light fermion~\cite{Affleck:1984fy}.  On the
other hand, the modulus will decay at tree level if $g\phi <
m_{\phi}$. Thus, the decay rate is given by:
\be
    \label{decay}
    \Gamma_{\phi} \approx \left\{
      \begin{array}{ll}
          {\frac{g^4 m_{\phi}^3}{\phi^2}} & (g\phi>m_{\phi}) \\
          [3mm]
          {\frac{g^2 m_{\phi}}{8\pi}} & (g\phi<m_{\phi})\ .
      \end{array}
      \right.
\ee
We are interested in the possibility that decay of the modulus $\phi$
is the final stage of reheating, so we assume that $\phi$ dominates
the energy density when it decays.\footnote{After all, if $\phi$ is
subdominant, its decay will just increase the energy density in the
dominant component.}  Then $H\sim m_{\phi}\phi/M_{P}$, so the
amplitude of the modulus and Hubble parameter during the decay are
given by \cite{Enqvist:2002rf,Enqvist:2003mr}
\be\label{Hd}
    \phi_d \approx \left\{
      \begin{array}{l}
          (g^4m_{\phi}^2M_{P})^{1/3} \\
          [3mm]
          g^2 M_{P}
      \end{array}
      \right.
\ ,\ \
    H_d \approx \left\{
      \begin{array}{l}
          {g^{4/3}
          \left(\frac{m_{\phi}}{M_{P}}\right)^{2/3}m_{\phi}} \\[3mm]
          g^2 m_{\phi}
      \end{array}
      \right.\ .
\ee
Depending on the mass of the modulus there are two distinct
possibilities. If the compactification has $m_{\phi}\sim \msm$ (which
is the natural region of parameter space to expect), the energy
density stored in the modulus at the time of decay is
$\rho_\phi\sim\msm^2\phi^2_{d}$. Assuming rapid thermalization of the
decay products, the reheat temperature is
\be\label{trhmodulus} T_{rh} \approx
0.1\sqrt{\Gamma_{d}M_{P}}\approx \left\{ \begin{array}{l}
0.1g^{2/3}\sqrt{\msm^{5/3}M_{P}^{1/3}} \\ 0.1g\sqrt{\msm
M_P}\end{array}\right.  \ .
\ee
In the first case (large $\phi_d$ and decay by loops), this
temperature is higher than $1/\lst\sim\msm$ as long as $g$ is not too
much less than 1, so the field theoretic approximation will break
down, and there will be some string theoretic phase. Then the final
stage of reheating will be given by the decay of long open strings
into radiation, as described in the next section.  In the second case,
we derive $g^3 M_P\lesssim \msm$ using (\ref{Hd}). This then implies
that the reheat temperature is above the string scale $\msm$ if the
coupling is in the range $\sqrt{\msm/M_P}\lesssim g\lesssim
(\msm/M_P)^{1/3}$ for which the final stage of reheating will also be
described by the decay of open strings.  If the coupling is $g\lesssim
\sqrt{\msm/M_P}$, the modulus will reheat the SM directly to
radiation.  In fact, since we expect a stringy phase when the density
is $\rho\gtrsim \msm^4$ or equivalently $H\gtrsim \msm^2/M_P$, the
Hubble parameter at modulus decay (\ref{Hd}) is exactly consistent
with this expectation.

When might the SM throat modulus come to dominate the energy density?
During the modulus evolution, its energy density should redshift no
more quickly than matter (it redshifts like matter during oscillation
and somewhat more slowly if it tracks a changing
potential). Therefore, we expect that the modulus will most likely
come to dominate the energy density if the other component is
radiation.  In other words, if the tunnelling rate (\ref{wdecay1}) is
low enough that the inflationary sector would reheat the SM throat
directly to radiation, the modulus could very well end up dominating
the energy density.  Another option is that the modulus could decay by
loops \textit{before} tunnelling occurs, in which case the
inflationary sector would still dominate the energy density, but the
modulus could reheat the SM sector to a long open string phase.


\subsubsection{Cosmological Consequences of the Modulus}\label{sss:cosmomod}

We will give a few examples of possible consequences of the SM throat
modulus decay (once again, we remind the reader that we do not have a
complete picture of the throat dynamics).

Clearly, one consequence is related to the fact that the roll of the
SM throat modulus is a necessarily higher dimensional effect.  In
terms of a simple Randall-Sundrum model, we would interpret it as the
roll of the radion field.  Since the 4D FRW equations are modified
when the radion is not stabilized, it would seem that this era may
have some form of nonstandard cosmology.

Additionally, the decay of the modulus might be responsible for
generating adiabatic density perturbatuions in the CMB
\cite{Lyth:2001nq,Enqvist:2002rf,Enqvist:2003mr,Enqvist:2004kg}.
Such a scalar can generate isocurvature fluctuations in the CMB
(although, since $m_\phi\sim H_\star$ during inflation, the large
scale fluctuations may be suppressed).  Let us consider what
happens if the modulus decays while the energy density of the
universe is dominated by an open string phase of density $\rho_o$.
This would lead to a partial conversion of the isocurvature
fluctuations to the adiabatic fluctuations and also a significant
nongaussianity parameter~\cite{Lyth:2001nq} \footnote{The
nongaussianity parameter is defined by the Bardeen potential which
arises while studying the temperature anisotropy
$\Phi=\Phi_{Gauss}+f_{NL}\Phi^2_{Gauss}$. In the curvaton scenario
$\Phi=-(r/5)(\delta\rho_{\phi}/\rho_{\phi})$. In the case of
modulus
$\delta\rho_{\phi}/\rho_{\phi}=2(\delta\phi/\phi)+(\delta\phi/\phi)^2$.
This leads to the non-Gaussianity parameter $f_{NL}\sim
5/4r$~\cite{Lyth:2001nq}.}
\be
f_{NL}\geq \frac{1}{r}\simeq \frac{1}{g^{4/3}}\left(
\frac{\mu^2}{\msm^4}\right)\left(\frac{\msm}{M_{P}}\right)^{2/3}
\ee
for the perturbations sourced by $\phi$.  Here, $r$ is the ratio
between the energy densities of the modulus and the strings at the
time of decay, $r=\msm^2\phi_d^2/\rho_o$.  We have assumed that the
modulus decays by loops before the strings convert to radiation, and
(to be conservative) we took the smallest value of $\rho_o\sim \mu^2$
for effective string tension $\mu$.  Obviously, once we assume that
the modulus obtains a Hubble-induced mass correction during inflation,
the perturbations of the modulus will be suppressed on large scales
during inflation and therefore the higher order perturbations as
well. The nongaussian signature may then be undetectable by the
forthcoming CMB experiments. A possible enhancement in non-Gaussianity
$f_{NL}\sim 50$ may arise naturally if the modulus mass is smaller
than the inflationary Hubble scale, $m_{\phi}\leq H_{inf}$, and it
decays non-perturbatively into the light degrees of
freedom~\cite{Enqvist:2004ey}.

In some models, the modulus decay could also directly produce
weakly interacting massive particles (WIMPs). For example, one of
the best known supersymmetric particles is the gravitino, which
the modulus could produce.  Then there are known cosmological
consequences.  For a TeV mass gravitino the decay time extends
through BBN, so therefore the initial abundance has to be small,
i.e., $n_{3}/s<10^{-10}$. The direct decay of the modulus into
gravitinos however generates significant abundance
\cite{Allahverdi:2004si}. This would then require heavy gravitinos
above $50$ TeV, which would decay before BBN. However, if the
gravitinos decay after the freeze out temperature of thermally
generated neutralinos, then their decay would also dilute the cold
dark matter abundance. Detailed model building issues have been
studied in \cite{Allahverdi:2005rh}.  The models we discuss might
avoid these problems entirely, also; the supersymmetry breaking
scale is not tied to the warping, so the gravitino could be
heavier than the modulus.

\section{Reheating due to open string decay}\label{s:decayreheat}

In this section, we will discuss reheating of SM modes (i.e., the
massless modes on the SM branes) through the decay of long open
strings on the SM branes.  Since we do not have a full understanding
of the SM throat geometry while $\msm < H<H_\star$, we pick up the
story when $H\sim \msm$, when we expect the SM throat to settle into
its true vacuum geometry, which we reviewed in section
\ref{ss:compact}.  Why bring in long open strings in this regime?  We
remind the reader that the effective field theory reheating
temperature (\ref{rhtemp}) can be sufficiently high that string modes
should have been excited during reheating, so there may be some
stringy phase of evolution.  In addition, even with $H\sim \msm$, the
4D energy density $\rho\sim \msm^2 M_P^2$ is much greater than the
effective string density $\rho_0\sim \msm^4\sim 1/\lst^4$.  Therefore,
we still expect a long string phase, and we will focus on fundamental
strings for simplicity.

Our goal is to establish a basic framework for the cosmology of open
strings, which we will see can be somewhat different than the standard
cosmology of (closed) cosmic strings.  This framework would apply in
many potential situations in string cosmology beyond our scenario ---
\textit{whenever} open strings are a significant source of energy
density.  In particular, we could imagine brane inflation with
low-energy supersymmetry, so that the standard model could live on
whatever branes survive the brane-antibrane annihilation.  In that
case, we might expect tachyon condensation to produce long open
strings on the surviving branes, which could provide at least some
component of reheating in the standard model.

To discuss reheating through long open strings, we first must describe
the distribution of open strings as well as the possible decay modes
and decay rates of the long strings.  Once we have reviewed that
physics, we will include the decay modes in the cosmological evolution
of the open string gas and discuss reheating from open strings to
massless SM modes.

\subsection{String Thermodynamics with D-branes}\label{ss:thermo}

In this section, we review essential features of the thermodynamics of
open strings, with more details in appendix \ref{ss:open}.  We
emphasize that this discussion is relevant for any discussion of open
strings in cosmology and not just our particular model.

Many authors have studied thermodynamics of open strings in the
background of D-branes \cite{Lee:1997iz,Abel:1999rq,Barbon:2004dd},
with the D-branes essentially considered as static objects.  Intrinsic
in our use of these results is the assumption that we can at least
approximately trust flat spacetime string thermodynamics in cosmology
when $H\lesssim 1/\lst$.  All the papers
\cite{Lee:1997iz,Abel:1999rq,Barbon:2004dd} agree that open strings
dominate closed strings if the number density of D-branes (in the
transverse volume) is greater than unity in string units, so the open
string thermodynamics is independent of closed string
thermodynamics. Therefore, we relegate a discussion of closed string
thermodynamics to appendix \ref{ss:closed}.  Although D-strings or
other wrapped D-branes might enter into thermal equilibrium at high
densities, we will leave that question for future work and focus on a
pure string gas for simplicity.\footnote{It is common string theory
folk-knowledge that a Hagedorn gas of strings undergoes a phase
transition to a black hole or black brane (i.e., something with a
horizon).  In appendix \ref{ss:vsbbrane}, we will argue that this is
not the case here, although we cannot rule out the appearance of a
horizon.}

One quantity we will need is the distribution of string number per
unit string length per volume parallel to the D-branes.  Specializing
to D3-branes, this is \cite{Lee:1997iz}
\be \label{opensoup} \mathcal{D}(\ell)=\frac{2aN_D^2}{b
V_\perp}e^{-\ell/L}\, ,\quad L=\frac{1}{\mu(\beta- \beta_H)}\, ,
\ee
where $L$ is the average length of the open strings, $\mu$ is the
string tension, and $\beta$ ($\beta_H$) is the inverse (Hagedorn)
temperature. In the above, both $a,b$ are proportional to $g_{s}$
(with a ratio $b/a\sim \mu$), $N_D$ is the number of branes, and
$V_\|,V_{\perp}$ are respectively the volumes along and transverse
to the D-branes. For now, we are working in 10D units for ease of
comparison to the literature, and throughout we take the
thermodynamic limit $E\propto V_\|\to \infty$.  Because it appears
in the distribution, $V_\perp$ is an important parameter in the
thermodynamics.  This volume is not the full volume of the
compactification manifold because the warp factor in the SM throat
acts as a potential for the compact position; rather we can treat
$V_\perp$ as the volume accessible to the strings through their
thermal and quantum fluctuations, as in
\cite{Jackson:2004zg}.\footnote{This procedure is somewhat ad hoc;
a proper treatment would include the potential in the statistical
description of the strings from the outset.}  It turns out that we
can approximate $V_\perp/(4\pi^2\ap)^3\sim 1$; the calculation is
given in detail in appendix \ref{ss:wsfluct}.

If we want to add D7-branes to the SM throat, (\ref{opensoup}) is
rearranged somewhat, and $a/b$ takes a different dimensionality.
However, in the end, the prefactor of the exponential takes the same
order of magnitude.  Also, each type of string ($33$, $77$, or $37$)
has its own distribution with $N_D^2\to N_{3}^2, N_{7}^2, N_3 N_7$
respectively.  In this paper, we will work mostly to order of
magnitude and simply write $N_D^2$.

So far, we have worked in ten dimensional units, but we can
translate to 4D effective units for the purpose of cosmology.
First, note that the 4D energy density is given by $\rho=\int
d\ell\,\mu\,\ell\,\mathcal{D}(\ell)$. Then using (\ref{opensoup}),
we can relate the average string length to this density
\be\label{avglengthopen} \bar\ell = e^{-\asm} L = \frac{1}{N_D}
\left(\frac{V_\perp}{(4\pi^2\ap)^3}\right)^{1/2}
\left(\frac{(2\pi\lst)^6\pi b}{a} \frac{\rho}{\msm^2}
\right)^{1/2}\approx \left(\frac{\rho}{N_D^2\msm^4}\right)^{1/2}
\lst\ ,\ee
where we have substituted $\mu=\msm^2/2\pi$, the effective SM
string tension. Note that the average string length is longer than
the string length for $\rho\gtrsim N_D^2\msm^4$, which is the
minimum density we consider.  Henceforth, all quantities are in 4D
units (i.e., measured in terms of the effective SM string scale)
except for $V_\perp$ and explicit factors of $\ap$. Let us also
make explicit that $\b\ell$ (and other lengths) are not the
lengths projected onto the 4D spacetime but rather the total
length in four-dimensional units. However, a simple random-walk
argument shows that $\b\ell$ and the projected length differ only
by factors of order unity.

Since we are also interested in the cosmological evolution of the open
string gas, we would like to know the equation of state.  Because the
compact directions are stabilized, we just need the pressure in the
noncompact directions, which is approximately
\be\label{pressure} p = \frac{1}{\beta} \left(\frac{4\pi a}{b}
\frac{N_D^2}{\msm^2} \frac{\rho}{V_\perp}\right)^{1/2}\approx
N_D\msm^2\sqrt{\rho} \ee
in the Hagedorn regime \cite{Abel:2002rs,Abel:2003jh,Cobas:2004dz}.
This equation of state is somewhat unusual in cosmology, but, in a
wide range of interest, we will see that the open strings behave like
a pressureless gas.

A critical question for cosmology is when the open strings are in
equilibrium.  The gas of strings can remain in thermodynamic
equilibrium as long as a typical string interacts once per Hubble time
(see the discussion of equilibration times in \cite{Lee:1997iz}).
Fortunately, the probability for two strings to interact given that
they cross is known to be \cite{Polchinski:1988cn,Jackson:2004zg}
\be\label{interconnectprob}
P = g_s^2\left[
\frac{(1-\cos\theta\sqrt{1-v^2})^2}{8 v\sin\theta\sqrt{1-v^2}} \right]
\frac{(4\pi^2\ap)^3}{V_\perp}\ .
\ee
Following typical arguments about cosmic strings \cite{Vilenkin},
we know that the empty area surrounding a string in a network of
density $\rho$ is $\mu/\rho$.  This leads to a crossing rate per
unit length of string of $v\rho/\mu$ and an interaction rate of
\be\label{rateperlength} \frac{d\Gamma}{d\ell} = g_s^2\left[
\frac{(1-\cos\theta\sqrt{1-v^2})^2}{8 \sin\theta\sqrt{1-v^2}} \right]
\frac{(4\pi^2\ap)^3}{V_\perp} \frac{\rho}{\mu}\ .  \ee
To get the interaction rate for a typical string, we just multiply by
the average string length as given above.  Then the equilibration rate
is
\be\label{avgrate} \Gamma \approx \frac{N_D^2 g_s^2}{\lst}
\left(\frac{\rho}{N_D^2\msm^4}\right)^{3/2} \ee
averaging over angles and velocities, in agreement with the rough
discussion of \cite{Lee:1997iz}, and again substituting
$\mu=\msm^2/2\pi$.  Of course, if the distribution of open strings
starts out far from equilibrium, we should use the typical length
scale of that distribution, rather than $\b\ell$, to determine the
equilibration rate.

Note that the dependence for (\ref{avgrate}) on the density of
open strings means that it is very difficult for the strings to
fall out of equilibrium.  Suppose that the open strings dominate
the universe, so $H\sim \sqrt{\rho}/M_P$.  Then a comparison gives
$\Gamma/H \sim N_D g_s^2 (M_P/\msm) (\rho/N_D^2\msm^4)$.  As long
as $\b\ell \gtrsim \lst$, the strings will remain in equilibrium.
In fact, it seems that the open strings will maintain equilibrium
unless they either start out very far from equilibrium or are
subdominant.

Finally, we should contrast these results to the usual intuition
developed from the study of cosmic strings, which is that long
cosmic strings enter a scaling solution rather than maintain
equilibrium (see \cite{Vilenkin} for example).  The difference
arises because of the different distribution of strings.  For
cosmic (closed) strings at typical densities, the gas of strings
is composed of very many small loops and a few very long strings.
As a consequence, two short loops cannot join to form a long
string, and, additionally, short strings will not often intersect
long strings.  However, long strings will intersect each other
roughly once a Hubble time, so they can decay by emission of short
loops relatively efficiently.  The long string population cannot
regain energy as quickly as it loses energy, so the strings fall
out of equilibrium.  In contrast, open strings are distributed
uniformly up to a length of roughly $\b\ell$, above which the
population decreases.  Therefore, there is a sizable population of
intermediate length strings, which can join with each other to
form long strings.  Therefore, the entire string gas can remain in
equilibrium if $\Gamma \gtrsim H$.

\subsection{Decay Rates of Long Open Strings}\label{ss:decayrates}

Since the reheating of the SM throat seems likely to lead to a
phase of long open strings, a crucial ingredient is the decay of
those strings.  Long open string decay has been the subject of
multiple studies, mostly in flat backgrounds with what we would
now recognize as D9-branes (or D25-branes in the bosonic case)
\cite{Mitchell:1988qe,Dai:1989cp,Okada:1989sd,Mitchell:1989uc}.
The relevant calculations of total decay rates for open strings on
superstring D-branes of arbitrary dimensionality appear in
\cite{Balasubramanian:1996xf}; we review the results here.  Even
though these decay rates are for static spacetimes, we will assume
their validity since the Hubble parameter is $H(t)\lesssim \msm$.
Additionally, these decay rates only apply, strictly speaking, to
leading Regge trajectory states, but we will use them for the
entire open string network.

Since the D3-branes do not fill the entire 10D spacetime, the decay
rate depends on the polarization of the open string in question,
longitudinal or transverse to the branes.  Realistically, there is a
wide variety (a ``discretuum,'' to borrow the term) of decay rates due
to the many possible polarizations of a long string, but we will lump
all the strings into two classes, those longitudinally polarized and
those transversely polarized.  In fact, these are precisely the two
cases considered by \cite{Balasubramanian:1996xf}.

The decay of a string polarized longitudinally to a D3-brane is
nearly a splitting rate per unit length.  If we restore factors of
$g_s$ and write $N$, the excitation level, in terms of $\ell$, the
string's length,\footnote{The conversion is $\sqrt{N}=\ell/4\lst$,
which follows from (for example) the light-cone Hamiltonian for a
string configuration $X^1=(\ell/2)\cos\sigma \cos\omega\tau$,
$X^2=(\ell/2)\cos\sigma \sin\omega\tau$, $0\leq\sigma\leq\pi$,
$\ell\omega = 2$.  This is explained in \cite{Dai:1989cp}.  The
calculation turns out to be the same for a transversely polarized
string rotating in two transverse directions.}
\cite{Balasubramanian:1996xf} give
\be\label{longsplit}
\Gamma_{\mathnormal{split}} = \frac{(2\pi)^3}{2\sqrt{2}} \frac{N_D
g_s}{\lst} \left(\frac{\ell}{\lst}\right)\left[\frac{1}{\pi}\ln\left(
\frac{\ell}{4\lst}\right)\right]^{-3}\ .
\ee
We have inserted an additional factor of $N_D$ compared to
\cite{Balasubramanian:1996xf} because the string can split on any of
the D3-branes.  Note that the log factors suppress this rate compared
to a decay rate per unit length; physically this means that the string
prefers slightly to split near its ends.  These logarithm factors are
caused by quantum mechanical fluctuations of the string away from the
worldvolume of the D3-brane.  For now, the important thing to note is
that even a longitudinally polarized string accesses the transverse
(compact) dimensions with a distance scale that grows logarithmically
in the length of the string.  Finally, we should note that splitting
has been included in the analysis of open string equilibrium, even
though it is not the dominant interaction (see \cite{Lee:1997iz}).

On the other hand, transversely polarized strings decay only by
radiating massless strings from their end points
\cite{Balasubramanian:1996xf} with a rate
\be\label{transradiate}
\Gamma_{\mathnormal{rad}} = \frac{\sqrt{2}(2\pi)^4}{8}
\frac{N_D g_s}{\ell}\ .\ee
Note that completely transversely polarized strings do not really
have any length in the noncompact dimensions. We assume,
though,that a typical ``transversely'' polarized string has about
the same length of string along the brane directions as it does
transverse to them.  This is the same assumption we made in the
thermodynamics. The reader should also be aware that this decay
rate for the leading Regge trajectory strings is slower than
expected for a typical transversely polarized string.  However, as
long as a string has any significant transverse polarization, it
will only decay to SM modes by radiation from its endpoints, and
we take the rate (\ref{transradiate}).

The key issue is to determine whether the long strings are
longitudinally or transversely polarized.  Let us quickly work through
the case of strings in thermal equilibrium near the Hagedorn
temperature.  Our rationale is as follows: if the linear scale of
quantum fluctuations of a longitudinal string is as great as the
linear scale of thermal fluctuations, then the strings can be
considered to be longitudinally polarized.  Otherwise, the strings are
transversely polarized.  Since the Hagedorn strings can always access
three of the dimensions of the SM throat tip, then we take the thermal
length scale to be $L/2\pi\sqrt{\ap} = (g_s n_R)^{1/2}/(4\pi)^{1/3}$,
as follows from our expression for $V_\perp$ (\ref{thermalvol}).  The
quantum length scale we can extract from the decay rate
(\ref{longsplit}) is $L/2\pi\sqrt{\ap} = \left[
\ln(\ell/4\lst)/\pi\right]^{1/2}$.  Therefore, we expect that open
strings will be longitudinally polarized only for lengths longer than
some critical value,
\be\label{longcondition}
\ell \gtrsim \ell_c\equiv 4\lst\, \exp\left[\left(\frac{\pi}{16}\right)^{1/3}
g_s n_R\right]\approx 10^3 \lst\ .\ee
The numerical value is given for $g_s n_R\sim 10$.  For strings at
equilibrium, there will only be many longitudinally polarized strings
if $\b\ell \gtrsim 10^3 \lst$, which would require
$\rho\gtrsim 10^6 N_D^2\msm^4$.  (We
should mention that different compactification models will give
different formulae for $\ell_c$.)

\subsection{Evolution of String Network and Reheating}\label{ss:network}

Here we will discuss the transfer of energy density from long open
strings into radiation.

\subsubsection{Strings in Equilibrium}

For specificity, we will begin with the case of equilibrium open
strings, giving a rough analysis of reheating into
radiation.\footnote{Throughout, we will assume that, by this stage
of reheating, the energy density in KK modes is subdominant. If
any KK modes are in equilibrium, they are a gas at temperature
$T\sim\msm$; nonequilibrium KK modes decay with a rate
$\dot\rho\propto \rho_{KK}$, which is much faster than \ref{loss2}
below for $\rho_{KK}\gtrsim\rho_o$ (unless the KK modes are nearly
stable).}

If we start by ignoring the decay of open strings to radiation, the
continuity equation, with the pressure (\ref{pressure}), is
\be\label{continuity1} \dot\rho +3\frac{\dot a}{a} \left(\rho
+N_D\msm^2 \rho^{1/2}\right)=0\ ,\ee
which has solution
\be\label{redshifting1} \rho = \left[ \left(\rho_0^{1/2}+
N_D\msm^2\right)\left(\frac{a_0}{a} \right)^{3/2}
-N_D\msm^2\right]^2\ .\ee
Here the subscript $0$ indicates an initial value.  Note that for
densities much larger than $N_D^2\msm^4$, the strings redshift as if
they are pressureless, which is not surprising given that the pressure
grows more slowly than linearly with density.  The deviation from this
matter-like redshifting is of order $N_D\msm^2/\rho^{1/2}$.  We should
remember, however, that, as this reaches order unity, the strings will
become an equilibrium bath of radiation.  Therefore, we will mostly
consider the open string gas to be pressureless.

Next we will consider the energy loss to radiation by transversely
polarized strings (remember that splitting of longitudinally polarized
strings is accounted for by the thermodynamics).  The rate of density
loss is simply written as
\be\label{loss1}
\dot\rho |_{\mathnormal{rad}} = \int_{\lst}^{\ell_c} d\ell\,
\mathcal{D}(\ell)\Gamma_{\mathnormal{rad}}(\ell) \Delta E(\ell) \ ,\ee
where $\Delta E(\ell)$ is the energy lost by a string of length $\ell$
in each radiative decay ($\mathcal{D}(\ell)$ and
$\Gamma_{\mathnormal{rad}}(\ell)$ are given respectively by
(\ref{opensoup}) and (\ref{transradiate})).  We can calculate $\Delta
E$ by noting that the radiation process decreases the level of a
transversely polarized string by $1$
\cite{Balasubramanian:1996xf}. Working in the center of mass frame,
$E^2 = N/\lst^2$, so
\be\label{deltaE}
\Delta E(\ell) = -\left(2\sqrt{N}\lst\right)^{-1} \approx 2/\ell\ ,
\ee
using our identification of $N=(\ell/4\lst)^2$.  This estimate assumes
that the string is long enough to neglect recoil and that relativistic
energy shifts of the emitted radiation averages out.  The lower limit
on the integral comes from insisting that the strings be long, and the
upper limit comes from the requirement that they be transversely
polarized.

Putting everything together, we find
\be\label{loss2} \dot\rho |_{\mathnormal{rad}}(\b\ell) \approx
-\frac{(N_D g_s)(N_D^2\msm^4)}{\b\ell} \Gamma(-1; \lst/\b\ell,
\ell_c/\b\ell)\approx \frac{(N_D g_s)(N_D^2\msm^4)}{\lst}\,, \ee
is given by an incomplete gamma function.  The final approximation is
appropriate for $\b\ell \gg \lst$, which we will assume
throughout. All told, the SM sector energy in radiation and open
strings obeys the equations
\bea \dot\rho_o +3\frac{\dot a}{a} \rho_o + \frac{(N_D g_s)(N_D^2
\msm^4)} {\lst} &=&0\,,\nonumber \\
\dot\rho_r + 4\frac{\dot a}{a} \rho_r - \frac{(N_D g_s)(N_D^2
\msm^4)}{\lst} &=& 0\ .
\label{coupledrho} \eea
These equations can be solved analytically; taking $A=2(N_D
g_s)(N_D^2\msm^4)/\lst$, we find
\bea \rho_o = -\frac{2 A}{9 H} +C_1\left(\frac{a_0}{a}\right)^3&,&
\rho_r = \frac{2 A}{11 H}+C_2 \left(\frac{a_0}{a}\right)^4\
\textnormal{for}\ \rho_o > \rho_r\label{matterdominated}\\
\rho_o = -\frac{A}{5 H} +D_1\left(\frac{a_0}{a}\right)^3&,& \rho_r
= \frac{A}{6 H}+D_2 \left(\frac{a_0}{a}\right)^4\
\textnormal{for}\ \rho_o < \rho_r\ ,
\label{raddominated}\eea
where $a_0$ is the initial scale factor for the given era.

We can use (\ref{matterdominated}) and (\ref{raddominated}) to get
a qualitative picture of reheating by open strings.  Imagine that
we take our initial conditions to be $H\sim \msm$ and $\rho_r$
negligible. Then the initial phase is dominated by the strings,
$C_1\sim M_P^2\msm^2$, and the equality between strings and
radiation occurs at about $H_{eq}\sim N_D g_s^{1/3}
(\msm/M_P)^{2/3} \msm$. Following into the radiation era,
reheating ends when $\rho_o\sim 0$, which turns out to be at
$H_f\lesssim H_{eq}$.  In the end, we find a final radiation
density of $\rho_r \lesssim (g_s M_P/\msm)^{2/3} N_D^2 \msm^4$.
This result is a bit larger than the typical string density
$\msm^4$, so a more detailed analysis of energy transfer between
strings and radiation is clearly necessary.  However, it seems
likely that the final reheat density and temperature for SM
radiation will be around $\rho\sim \msm^4$ and $T\sim \msm$

respectively.\footnote{As $\msm$ is a free parameter in our
discussion, we note that the above result raises various
phenomenological challenges if $\msm$ is too low. If $\msm\sim
{\cal O}(1~{\rm TeV})$ then the most important problem will be to
obtain the observed baryonic asymmetry~$n_{B}/n_{\gamma}\sim
6.1\times 10^{-10}$~\cite{Bennett:2003ba}, because of the larger
parameter space, electroweak baryogenesis in minimal
supersymmetric SM in principle has a much better chance to
succeed. However, there are a number of important constraints, and
lately Higgs searches at LEP have narrowed down the parameter
space to the point where it has all but
disappeared~\cite{Cline:2000nw,Carena:2002es}. However for an
intermediate scale $\msm$ baryogenesis could take place via other
means~\cite{Enqvist:2003gh}.  The other challenge will be to
obtain the right abundance for the cold dark matter. Below
$\msm\sim 1$~TeV the only viable candidate is the coherent
oscillations of the axionic cold dark
matter~\cite{Preskill:1982cy,Abbott:1982af,Dine:1982ah}. For an
intermediate string scale it is possible to have a viable KK cold
dark matter whose abundance could be either obtained
non-perturbatively from the decay of the modulus, or from thermal
excitations.}


\subsubsection{Nonequilibrium Strings}\label{sss:noneq}

In this section, we will give a few comments about the evolution of
nonequilibrium strings, for which we do not a priori know the
distribution of lengths.  As we have seen in the equilibrium case, the
string distribution was important in the rate of decay to radiation;
we suspect that a nonequilibrium string distribution would be just as
important. With that in mind, we can give only a few general comments.

Out of equilibrium, we must distinguish two populations of open
strings, superhorizon ($\ell>1/H$) and subhorizon ($\ell<1/H$).  While
superhorizon strings grow with the cosmological expansion, the
subhorizon strings can only decay.  In addition, since the strings
interact rarely, there is an effective maximum to the length of any
given string, due to the splitting of longitudinal polarizations.
That is, as any string approaches $\ell_c$ defined in
(\ref{longcondition}), the SM branes come within the quantum radius of
the string, so the string can split with a string-scale rate
\textit{per unit length} (treating the logarithm as roughly constant
near $\ell_c$).  Therefore, the longitudinal strings rapidly break
into smaller strings, so there should be a large population of
superhorizon strings only while $\ell_c$ is larger than the Hubble
radius.  (There may be a few strings with atypically large deviations
from the SM branes, and these few strings could be transversely
polarized but longer than $\ell_c$.)

Let us consider the superhorizon strings a little more carefully.
Since the open strings are out of equilibrium, they are not
interacting even once a Hubble time, so they cannot enter a scaling
solution.  In that case, the superhorizon strings could enter a
string-like phase, redshifting as $a^{-2}$, at least for
$\ell_c\gtrsim 1/H$.  Then, as a given string stretches to a length of
$\ell_c$, it will split, so the distribution of strings will evolve;
numerical simulations will be necessary to understand the evolving
distribution.

Finally, we note that nearly all the nonequilibrium strings should
decay to radiation on a time scale given roughly by $\ell_c$ for two
reasons.  First, the radiative decay rate (\ref{transradiate}), for
typical strings, is roughly $\Gamma\gtrsim 1/\ell_c$.  Second, by
$t\sim \ell_c$, typical transversely polarized strings will be
subhorizon length, so they will no longer stretch and will simply
decay.  Only a few, atypically long transversely polarized strings
should survive beyond $t\sim \ell_c$.  Additionally, these should
eventually become subhorizon or longitudinally polarized and decay as
well.  Since $H\sim 1/\ell_c$ occurs much before nucleosynthesis, it
seems that light element abundances are protected in this case.

\section{Discussion: Possible Signatures}\label{s:discuss}

Finally, in this section, we will speculate on the possible
cosmological signatures of the light SM sector strings in these warped
models.  The discussion will be ordered chronologically, from
inflation onward.

\subsection{Fluctuations in Inflation}\label{ss:fluctuations}

As we mentioned in the introduction, high energy physics can alter
the inflaton fluctuation spectrum at the earliest stages of
inflation
\cite{Kaloper:2002uj,Burgess:2003zw,Schalm:2004qk,Porrati:2004gz,
Greene:2004np,Easther:2004vq,Collins:2005nu,Easther:2005yr,Collins:2005cm,
Martin:2000xs,Brandenberger:2000wr,
Martin:2000bv,Brandenberger:2002hs,Martin:2003kp,Brandenberger:2004kx,
Niemeyer:2000eh,Niemeyer:2001qe,Kempf:2001fa,Niemeyer:2002kh,Easther:2001fi,
Easther:2001fz,Easther:2002xe,Danielsson:2004xw,Burgess:2002ub,
Danielsson:2002kx,Danielsson:2002qh}.
So far, in the absence of a theory of quantum gravity or a
specific realization of inflation in string theory, the effects of
``trans-Planckian'' physics on the inflaton have been modeled by
modified high-energy dispersion relations, modified initial
conditions (vacua) for the inflaton, and irrelevant operators in
the inflation EFT.  The key point is that, depending on the type
of signals expected, $H/M\gtrsim 1/10$ (or $H^2/M^2\gtrsim 1/10$)
are necessary for corrections to the usual spectrum to be
observable \cite{Kaloper:2002uj}. Here, we have argued that a
significantly warped SM throat will naturally provide
$H/M\sim\mathcal{O}(10^{-2}-1)$ during inflation
--- see equation (\ref{inflatestring}). Here $M$ may
represent either the effective Kaluza-Klein and string scales in
the SM throat. Hence it is likely that higher-dimensional and
stringy physics may be observable.

In fact, we can estimate the strength of the corrections.  One of the
(leading) $\mathcal{O}(\ap)^3$ corrections to the supergravity action
\cite{Grisaru:1986vi,Myers:1987qx} contains a term (up to prefactors)
\be\label{riemann4}
R^{\alpha\beta\mu\nu}R_{\lambda\rho\mu\nu}R_\alpha{}^{mn\lambda}
R^\rho{}_{mn\beta}\approx R^{\mu\nu\lambda\rho}R_{\mu\nu\lambda\rho}
\left(\Del_m \Del_n A\right)^2\ ,\ee
where the right hand side follows by conformally rescaling the 10D metric.
Furthermore, $R^{\mu\nu\lambda\rho}R_{\mu\nu\lambda\rho}\sim e^{-4A}H^4$
(plus terms depending on derivatives of the warp factor)
by counting powers of the warp factor.
If we dimensionally reduce this term, the contribution to the action is
\bea
\delta S &\approx& M_{10}^8\ap{}^3 \int d^4x\, \sqrt{-g_4} \int d^6x\,
\sqrt{g_6}\, H^4 \left(\Del_m \Del_n A\right)^2\nonumber\\
&\approx& M_P^2 \ap{}^3 H^4 \int d^4x\, \sqrt{-g_4} \left\langle
e^{-2A} \left(\Del_m \Del_n A\right)^2\right\rangle\
,\label{deltaS} \eea
where 
\be\label{avg}
\langle\cdots\rangle\equiv\frac{1}{V_w}\int d^6x \sqrt{g_6}\,e^{2A}\cdots =
\frac{1}{V_w}\int d^6x \sqrt{\hat g_6}\, e^{-4A}\cdots
\ee 
is the average over the compact
manifold, weighted by the same power of the warp factor that gives the
warped volume.  Since the warp factor is very
small near the tip of the SM throat, this region dominates the
average.  Therefore, we evaluate the average there; by using
Riemann normal coordinates, we can estimate $\left(\Del_m \Del_n
A\right)^2\sim (g_s n_R\ap)^{-2}$. In the end, we find
\be\label{deltaS2} \delta S\approx \frac{(M_P H)^2}{(g_s n_R)^2}
\frac{H^2}{\msm^2}\ .\ee
This is suppressed only by $(g_s n_R\ap)^{-2} (H/\msm)^2$ compared
to the Einstein-Hilbert term, so we expect that it may be
detectable.

Distinguishing our string compactification effects from field
theoretic models will likely be difficult, however.  One
suggestion is to examine, order by order in $H/M$, the higher
dimension operators generated by string theory and KK physics. If
some characteristic pattern of coefficients for these
higher-derivative terms emerges (or better yet, if some of the
coefficients are forced to vanish by a symmetry), this pattern may
be used to single out compactification physics as the source of
any signal. The work of \cite{Hwang:2005hb} regarding inflaton
fluctuations in higher-derivative gravity may be useful for
determining the actual CMB spectrum in such a case. In a similar
spirit, \cite{Silverstein:2003hf,Alishahiha:2004eh,Chen:2005ad} have given a
characteristic pattern of modifications to the CMB that might
arise in brane inflation due to noncanonical inflaton kinetic
terms.  For another approach, relying on post-inflation evolution,
see \cite{Hannestad:2004ts}.

One should also consider the possibility of inherently stringy
effects during inflation. While the light SM sector strings have
$\lst$ less than the Hubble length $1/H$, we do not expect the
separation of scales to be very large. Hence, for example, the
creation of long strings, as discussed at the end of section
\ref{ss:problems}, is suppressed but not enormously. So one might
find a moderate production of strings during inflation. There
might also be unusual quantum effects coming from macroscopic
virtual strings during inflation or fluctuations of the string
field. Again the point is that the separation of $\lst$ and $1/H$
should not be immense. So perhaps virtual processes can modify the
quantum behavior of the inflaton with nonlocal or stringy effects.
While necessarily vague, it may be that string field theory will
be necessary to develop these ideas.

\subsection{Non-Standard Cosmology During Reheating}\label{ss:nonstandard}

We also noted that, for $H_\star \gtrsim H(t)\gtrsim\msm$, the SM
throat geometry will evolve from its inflationary state to the
true ground state geometry.  In section \ref{ss:modulus}, we
modeled this evolution in terms of EFT by discussing the
qualitative behavior of the SM throat deformation modulus.  It is
worth reiterating the point that, in a Randall-Sundrum model of
the compactification, the evolution of the SM throat would be
modeled by a rolling of the radion.  What changes to standard FRW
cosmology might result is an interesting question for future work;
it seems likely that there would be some corrections for the
following reason.  As explained in
\cite{Binetruy:1999ut,Csaki:1999mp}, the usual Friedmann equation
is modified in a Randall-Sundrum universe if the radion is not
stabilized; essentially, this fact is because the Einstein
equations are five-dimensional.  Therefore, we might guess that,
if the radion is rolling in its potential (that is, it is not
completely stabilized by its potential) due to some finite energy
density, the Friedmann equation will be corrected. Additionally,
in solving the higher dimensional equations of motion, the space
and time directions may develop different warp factors, leading to
a violation of Lorentz invariance (see the solutions of
\cite{Lesgourgues:2000tj,Kanno:2004yb} for example).

\subsection{Early Structure Formation (Primordial Black Holes)}\label{ss:bh}

We have argued that the universe is dominated by a gas of long open
strings with negligible pressure after the end of inflation.
Therefore, the curvature fluctuations created during inflation enter
the Hubble radius and perturb the long open strings.  As is well
known, inside the Hubble radius, the linear perturbations in cold dark
matter grow, see \cite{Kolb:1990vq}:
\be
\delta(t)=\delta(t_{i})\left(\frac{t}{t_{i}}\right)^{2/3}\, ,
\ee
where $\delta$ is the density contrast and the time $t_{i}$ is
chosen to normalize the perturbations entering the Hubble radius.
This is analogous to the Jean's instability in a static universe,
but expansion slows the exponential growth in the density contrast
to a power law.  The overdense regions expand slowly, reach a
maximum radius, contract, and eventually virialize to form a bound
nonlinear system with density contrast of $\delta \sim
\mathcal{O}(1)$.

There are many possibilities for these gravitationaly bound
systems which one can entertain, such as objects made up of SM
baryons, the earliest stars. Nevertheless, as the long strings
decay into radiation, the most tenacious objects would be, of
course, black holes (which might be the only structure to
survive). These primordial black holes \cite{Carr:1974nx} can
survive until the present, unless their Hawking radiation is
important. The lifetime of an evaporating black hole is given by
\be \frac{\tau}{10^{17}\,\textnormal{sec}}\simeq \left(\frac{M}{10^{15}\,
\textnormal{grams}}\right)^3\,.
\ee
Note that a black hole of initial mass $M\sim 10^{15}\,
\textnormal{g}$ will evaporate at the present epoch while $M\sim
10^{9}\,\textnormal{g}$ will evaporate around the time of
nucleosynthesis.  Assuming that we form black holes with
approximately the mass contained in the horizon (for a review
see~\cite{Liddle:1998nt}), we have
\be M_{\textnormal{\scriptsize HOR}}\approx 10^{9}\,\textnormal{g}
\left(\frac{(10^{11}\,\textnormal{GeV})^4}
{\rho_{o}}\right)^{1/2}\ \ \textnormal{for}\ \
\rho_{o}>N_{D}^2\msm^4\,. \ee
If the string dominated phase lasts until $\rho_{o}\sim
(10^{8}\,\textnormal{GeV})^4$, black holes with mass $M\sim
10^{15}\,\textnormal{g}$ are formed and can live through today. In
any case, massive black holes above $10^{9}\,\textnormal{g}$ are
astrophysically interesting. Depending on their present
abundances, they can be safe with a distinct astrophysical
signatures. For example, black holes with mass $M\sim 5\times
10^{14}\,\textnormal{g}$ are already constrained from the current
$\gamma$-ray bursts as long as the fraction of black hole energy
density to the energy density of everything else is
$\alpha_{evap}<10^{-26}M/M_{P}$ \cite{Halzen:1991uw}.  In the
extreme case that $\msm\sim 1\,\textnormal{TeV}$, the string
dominated phase lasts until the electroweak phase, and black holes
of mass $M\sim 10^{25}\,\textnormal{g}$ can be formed. They would
live through the present; it would be interesting to see if there
are observational constraints on such black holes.

\subsection{Gravitational Waves}\label{ss:gravwaves}

There are also several possibilities for the generation of
gravitational waves (beyond those created as fluctuations during
inflation).

One possibility occurs because the transition from inflation to a
long string dominated phase is not instantaneous. There is a brief
period of radiation which however quickly becomes subdominant.
During these transitions there would be a slight bump in the
gravitational wave spectrum which would have a frequency cut-off
roughly given by the geometric mean of two scales:
$\omega=2\pi(a(tr)H_{tr}/a(\tau)\approx 2\pi\sqrt{H_{0}H_{tr}}$,
possibly allowing the detectability of such gravitational waves by
future experiments.

Another possibility is that the long strings themselves source
gravitational waves.  Clearly, during the long open string phase,
there will be a subdominant long closed string component (see
appendix \ref{ss:closed}), and these closed strings will
eventually decay into gravitons.  Additionally, long open strings
can dynamically develop cusps, just as closed strings, which then
produce beams of gravity waves (see \cite{Vilenkin} for the
appropriate cosmic string literature).

\subsection{Other Effects}\label{ss:othersigs}

The prolonged phase of matter (open string) domination has other mild
effects in cosmology. For instance the required number of e-foldings
during inflation could be considerably smaller by:
\be N(k)=62-\frac{1}{12}\ln\frac{V_{end}}{N_{D}^2\msm^4}\,. \ee
For $V_{end}\sim 10^{64}~({\rm GeV})^4$ and $N_{D}^2\msm^4\sim
10^{12}~({\rm GeV})^4$, the number of e-foldings required to
expalin the observed CMB spectrum and the flatness and the
homogeneity problem is only $52$ e-foldings.

Additionally, as \cite{Kofman:2005yz} has pointed out, KK modes
with large compact angular momentum in warped throats can have
very long lifetimes because light (and SM) modes do not carry the
same angular momentum on the internal space. Since either
reheating from the inflationary sector or the SM throat modulus
can populate these KK modes, we have a natural mechanism for
production of relics.  What remains to be determined is whether
and at what density these KK relics freeze out.  Another possible 
candidate relic is a leading Regge trajectory open string polarized
transverse to the SM brane.  However, since the open string gas stays in
equilibrium very easily, we would need some nonthermal production process
for these strings.

Finally, we mention that the long open string gas near the
Hagedorn temperature suffers from large thermal fluctuations (in
the canonical ensemble).\footnote{Independently,
\cite{Brandenberger} are studying thermal fluctuations of a string
ensemble in the context of noninflationary cosmology.  We thank R.
Brandenberger for bringing their work to our attention.}
Therefore, the density and temperature should vary from Hubble
region to Hubble region (and even within Hubble regions).  If
these fluctuations can survive being washed out by the subsequent
radiation phase, they could leave a significant imprint on the
CMB.

\section{Summary}\label{s:conclusion}

To summarize, we have considered string inflation in the context
of warped compactifications with multiple throats. In scenarios
where the warping differs significantly between throats (such as
in scenarios where the inflationary scale is much higher than the
SM scale), the standard approach of treating inflation as a
perturbation on the ground state configuration is inconsistent.
Not only would the 4D EFT break down due to KK mode excitations,
but the supergravity approximation would also due to large 10D
curvatures. Instead, we argued that the ``vacuum'' geometry of
highly warped regions (roughly $e^{\asm}\lesssim e^{2A_{inf}}$ ---
see (\ref{curve})) is modified during inflation.  Our argument was
based on intuition from 4D EFT, and we looked for a consistent
picture of the physics. As it turned out, the string (and KK) mass
scale is just above the Hubble scale, or $\lst H\lesssim 1$, in
the modified geometry. Effective field theory is then (just)
consistent, but there may be corrections from KK and string
physics, which can lead to potentially observable perturbations of
the CMB.

Also, we analyzed reheating of the SM sector, following the
``warped reheating'' scenario of \cite{Barnaby:2004gg}. We also
stress that we are taking the view that the compactification is
never in its vacuum until well after inflation ends. The energy
stored in the SM throat itself then and the relaxation of the
geometry to its ``ground state'' may have interesting implications
for reheating, as discussed in section \ref{s:warpreheat}B. In
both cases, we saw that there is a strong possibility for
reheating to create a phase of long open strings in the SM throat.
We provided a rough analysis of the open string phase based on
thermal equilibrium, finding that the strings redshift like
matter. Additionally, the strings slowly decay to SM radiation,
eventually leaving a radiation bath at about the Hagedorn
temperature $\msm$. Standard hot big bang cosmology can proceed as
normal from that point.

Finally, in the previous section, we listed some possible
observational consequences and directions for future research.  It
will be very important to analyze these possible signals, especially
to see if there is some characteristic signature of string theory or
KK physics, nongaussianity and inhomogeneous reheating due to strings,
generating CMB anisotropy due to string fluctuations during the
hagedorn phase, relic KK particles,
and the gravity wave signals from various transitions
in the early Universe.

Certainly our preliminary investigation leaves many open
questions. The most pressing amongst these is perhaps to find a
more precise description of the SM throat geometry during
inflation. While this is undoubtedly a technically challenging
question, our arguments indicate that it should be within the
reach of supergravity calculations. In any event, the issues
raised here should be a strong precaution for the ``modular''
approach to model building where separate throats are introduced
to produce apparently separate scales for, e.g., inflation, SM
physics and the cosmological constant.

\begin{acknowledgments}

We would like to thank R. Brandenberger, A. Buchel, C. Burgess, X.-G. Chen,
S. Hofmann, S. Kachru, L. Kofman, J. Polchinski, M. Savage, and M. Schulz for
useful comments and conversations.  The work of ARF has been
supported by a John A. McCone Fellowship in Theoretical Physics at
the California Institute of Technology. Research at the Perimeter
Institute is supported in part by funds from NSERC of Canada and
MEDT of Ontario. RCM is further supported by an NSERC Discovery
grant.  ARF and AM would also like to thank the Aspen Center for Physics
for hospitality during the final stages of this work.

\end{acknowledgments}


\appendix

\section{Conventions and Notation}\label{s:conventions}

Here we will describe conventions and notation used throughout the
paper.

We start by describing the relation of the various fundamental scales
of the compactification.  The 10D string frame gravitational action is
\be\label{action10d}
S = \frac{1}{2(2\pi)^7\ap{}^4} \int d^{10}x\sqrt{-g} e^{-2\varphi} R+\cdots\ ,
\ee
where $\varphi$ is the dilaton.  Since we assume throughout the
paper that the dilaton (as well as the volume modulus of the
compactification) is stabilized at a high energy scale, we
separate it into a VEV and a fluctuation, $\varphi=\ln g_s
+\delta\varphi$.  Then the background value of the
Einstein-Hilbert term prefactor is
$M_{10}^8/2=1/2(2\pi)^7\ap{}^4g_s^2$.  The 10D Planck constant is
slightly larger than the fundamental string scale in the
perturbative limit, i.e., $\ap M^2_{10}\gtrsim 1$.

In the dimensional reduction, the zero mode $h_{\mu\nu}$ of the
graviton has a wavefunction given by the warp factor, $\delta
g_{\mu\nu} = e^{2A}h_{\mu\nu}$, where $\delta g_{\mu\nu}$ is the
perturbation of the 10D metric (\ref{warpmetric}) with
polarization in the 4D spacetime. Therefore, scaling out the warp
factor, we find that the action is
\be\label{action4d1}
S=\frac{M_{10}^8 V_w}{2}\int d^4x\sqrt{-g_4} \left(1+\frac{\delta V_w}{V_w}
\right)e^{-2\delta\varphi}R(g_4)+\cdots\ ,
\ee
where the warped volume is
\be\label{warpedV}
V_w \equiv \left(2\pi\sqrt{\ap}\right)^6 e^{6u}
\equiv\int d^6x\sqrt{g_6} e^{2A} = \int d^6x\sqrt{\hat g} e^{-4A}
\ee
and $\hat g_{mn}$ is the underlying Calabi-Yau metric.  Equation
(\ref{warpedV}) then defines the volume modulus $u$.  Since
$e^{-4A}$ grows only as $1/r^4$ as $r\to 0$ in an AdS throat (and
less rapidly in other throats), we can see that $V_w\sim
V_{\mathnormal{CY}}$ is approximately the volume of the Calabi-Yau
manifold.  Therefore, $\sqrt{\ap} e^u$ is approximately the linear
scale of the Calabi-Yau and gives the KK mass scale for modes not
localized in highly warped regions.  If we henceforth refer to $u$
as the expectation value and $\delta u$ as the fluctuation, the 4D
Planck constant becomes
\be\label{planck4d}
M_P^2 = M_{10}^8 \left(2\pi\sqrt{\ap}\right)^6 e^{6u}
= \frac{e^{6u}}{2\pi\ap g_s^2}\ ,\ee
and the graviton and scalar modes are decoupled by going to Einstein
frame $g^E_{\mu\nu}= e^{6\delta u-2\delta\varphi}g^4_{\mu\nu}$.  Note
that, since $\delta u,\delta\varphi$ are the (massive) fluctuations
only, there is no rescaling of masses from string to Einstein frame.

Since the supergravity 3-form fluxes play an important role in
stabilizing moduli and determining the geometry of warped regions of
the compactification, we will also define the flux quantum numbers
$n_{f,R,NS}$ used in the text.  Both the Ramond-Ramond (R) and
Neveu-Schwarz--Neveu-Schwarz (NS) fluxes (respectively $F$ and $H$)
satisfy a quantization condition when integrated on any 3-cycle $c$:
\be\label{fluxquantize}
\int_c F = 4\pi^2\ap n^c_R\ ,\ \int_c H = 4\pi^2\ap n^c_{NS}\ ,\
n^c_{R,NS}\in \mathbf{Z}\ .\ee
The fluxes are completely specified by the integers $n^c_{R,NS}$.
Ignoring warping, the fluxes induce masses for the complex
structure moduli of the Calabi-Yau as well as the dilaton.  These
masses are given (up to numerical constants) by
\cite{deAlwis:2003sn,Buchel:2003js,deAlwis:2004qh,Giddings:2005ff},
\be\label{fluxmass}
m^2\approx g^2_s \left|F-\left(C+\frac{i}{g_s}\right)H\right|^2 \approx
\frac{g_s^2}{\ap} e^{-6u}\sum_c \left| a_c \left(n_R^c -\left(C+
\frac{i}{g_s}\right)n_{NS}^c\right)\right|^2\equiv
\frac{g_s^2 n_f^2}{\ap}e^{-6u}\ ,\ee
where the sum runs over all the relevant 3-cycles and $a_c$ is a
numerical constant relating the components of the fluxes to the
integer $n^c_{R,NS}$.  Here $C$ is the scalar from the Ramond-Ramond
sector of the supergravity, which is also stabilized by the flux.

Since the fluxes also source the warp factor, they determine much
of the geometry of the warped throats.  There are two important
points for us, which we will illustrate in the case of the
deformed conifold throats of
\cite{Klebanov:2000nc,Klebanov:2000hb}.  The first point is that
the fluxes control the size of the tip of the throat.  For
example, the tip of a deformed conifold throat is
$\mathbf{R}^3\times S^3$, and the radius of the sphere is given by
the flux.  Taking $C=0$ and the Ramond-Ramond flux to wrap the
$S^3$, the radius of the sphere is roughly $R_{S^3}\sim\sqrt{g_s
n_R^{S^3} \ap}$.  (Henceforth, and in the rest of the paper, we
drop the superscript $S^3$ when we discuss a particular throat.
In this case, the Neveu-Schwarz--Neveu-Scharz flux wraps the cycle
dual to the $S^3$.)  The other point is that the fluxes determine
the warp factor in the throat.  For one, they actually fix the
value of the warp factor at the bottom of the throat, although the
details are unimportant to us.  In addition, the warp factor
behaves like the warp factor for AdS with a varying AdS radius,
and the fluxes control the AdS radius.  At the tip of the deformed
conifold throat, for example, the AdS radius is
$R_{\textnormal{\scriptsize AdS}}\sim \sqrt{g_s n_R\ap}$, making
the same assumptions about the flux and $C=0$ as above.  Whenever
we discuss $n_{R,NS}$ in this paper, we are referring to this
choice of flux and $C=0$ in the throat.

Finally, let us comment on the possible effect of the warp factor
on the moduli masses.  The effect of the warp factor should only
be important for moduli with wavefunctions localized in a throat
(or other significantly warped region), in which case we expect
that we would replace $\ap\to \lst^2$ (for the SM throat, and so
on) in equation (\ref{fluxmass}).  However, there may be
additional corrections due to derivatives of the warp factor.  In
the case of the deformation modulus of the SM throat, it is
possible that the dominant term may be
\be\label{fluxmasswarp}
m^2\sim \frac{g_s^2 n_f^2}{\lst^2}
\frac{1}{(R^2_{\textnormal{\scriptsize AdS}}/\ap)^p}=
\frac{g_s^2 n_f^2}{\lst^2}\frac{1}{(g_s n_R)^p}\ \ \textnormal{rather than}
\ \ m^2\sim\frac{g_s^2 n_f^2}{\lst^2}e^{-6u}
\ee
for some power $p$.  In the text, however, we will assume the
suppression by $e^{-6u}$ compared to the string scale, since the
dimensional reduction in the presence of warping is not completely
understood.


\section{Alternative Proposals for SM Throat}\label{s:alternatives}

Here we will give a few comments about two possible alternative
geometries to that discussed section \ref{s:stringinflate}. These
are modifications to the 10D geometry which might seem natural for
the SM throat during inflation.  We will also present arguments
why these modifications are not applicable in the present context
of interest.

\subsection{Black-brane Horizon}\label{sss:blackbrane}

The first alternative we consider is that a black-brane horizon
replaces the entire lower portion of the throat.  There is a
simple thermal reasoning for this picture; it is well-known that
any observer in de Sitter (or inflationary) spacetime sees a heat
bath at the Gibbons-Hawking temperature $T=H/2\pi$.  Then we might
imagine that the SM is just a gauge theory in a heat bath;
string-gauge duality would then suggest representing the thermal
gauge theory as a black hole with matching Hawking temperature in
the SM throat \cite{Witten:1998zw}.  In such a scenario during
inflation, the entire lower portion of the SM throat (including
the SM branes) are hidden by a horizon appearing in the throat. In
this case, the observable physics is then cut off at some warp
factor much large than the vacuum value $e^{\asm}$ --- so in this
respect it is similar to the modified geometry in section
\ref{s:stringinflate}. In the black horizon picture, the standard
model degrees of freedom are no longer open strings on D-branes
but rather closed strings near the horizon. At the end of
inflation, the horizon shrinks, the D-branes appear, and the SM
strings become open strings on the branes again, which should be
excited up to the energy of the black hole. This process can be
understood as a phase transition
\cite{Buchel:2000ch,Buchel:2001gw,Buchel:2001qi,Gubser:2001ri,
PandoZayas:2003jr,Aharony:2003sx}.  In fact, this type of
transition between thermal strings and black holes has been
advocated on general grounds \cite{Horowitz:1997nw}.

Black holes in these throats have been discussed in
\cite{Gubser:2001ri}, in which it was shown that the geometry
approximates a nonextremal black 3-brane at high temperature.  The
metric for the black 3-brane is
\cite{Horowitz:1991cd,Gubser:1996de}
\be
\label{black3}
ds^2=e^{2A(r)}\left[ -g(r)dt^2+dx^i dx^i\right] +e^{-2A(r)}\left[
g^{-1}(r)dr^2 +r^2ds_5^2\right]\ ,
\ee
where $ds_5^2$ is some compact five-dimensional Einstein metric.
For simplicity, take the six-dimensional throat to be
asymptotically a conifold (up to the warping and the fact that it
glues onto some compact space), so $ds_5^2$ is the metric of
$T^{1,1}$.  By setting the brane Hawking temperature to the de
Sitter temperature, we can approximate the effective string
tension for the SM throat using the warp factor $A_h$ \textit{at
the horizon}. After some algebra, we find an effective horizon
string length
\be
\label{horizonlength}
\lsh = \left(\frac{64}{g_s q}\right)^{1/4}\frac{1}{H}\ .
\ee
We should note that the D3-brane charge $q$ is an ``effective''
charge; the 5-form field strength varies over the conifold throat due
to the presence of 3-forms \cite{Klebanov:2000nc,Klebanov:2000hb}.
Following the considerations of \cite{Giddings:2001yu}, the ratio of
scales (warp factors) between the tip and horizon is
\be\label{ratio}
q = N_D+\frac{3g_s n_R^2}{2\pi} \ln\left(\lst/\lsh\right)
\ee
(in agreement with \cite{Gubser:2001ri}).  We can solve the
transcendental relation (\ref{horizonlength}) for $\lsh$ approximately
by substituting $H$ on the right hand side and ignoring the number
$N_D$ of SM branes.  If we assume $\msm\sim 1\,\textnormal{TeV}$ and
$H\sim 10^{13}\,\textnormal{GeV}$, we get $g_s q\sim 1000$ and
$\lsh\sim 1/2H$.  It is amusing to note that we find $\ell_h H\lesssim
1$ so simply using thermal physics.

We can now explain a clear problem with this alternative proposal.
The black-horizon, because it is thermal, carries an energy
density in the $\N=1$ 4D field theory which is dual to the throat
geometry (as per the AdS/CFT correspondence). In fact, following
\cite{Gubser:1996de}, we can argue that the thermal energy density
is
\be\label{densitybh2} \rho_{bb} = \frac{3\pi^4 \omega_5}{32} q^2 T^4 =
\frac{3\omega_5}{512}q^2 H^4\ ,
\ee
where $\omega_5$ is the volume of the angular part of the
compactification (which is a $T^{1,1}$ space here).  Furthermore, the
black brane horizon has the equation of
state of radiation, as follows from its entropy (see
\cite{Horowitz:1997nw}).  Now suppose we compare to the energy of the
dual gauge field theory in a de Sitter background.  Certainly, any
observer sees radiation of temperature $T=H/2\pi$, but this temperature
is invisible in any global sense.  Specifically, in any de Sitter-invariant
vacuum, the stress tensor is that of a cosmological constant
$\langle T_{\mu\nu}\rangle \propto g_{\mu\nu}$.  In fact, the correct
dual geometries for $\N=1$
gauge theories on de Sitter (with nondynamical gravity)
do not have thermal brane horizons, and they are reviewed in
section \ref{sss:bucheltseytlin} below.

Another point to note is that the energy density
(\ref{densitybh2}) grows as the square of the effective D3-brane
charge.  This density might seem consistent with the dual picture
of a gauge theory during inflation (in fact, the factor of $q^2$
is consistent with the number of adjoint fields); however,
determining the energy density of the gauge fields is more subtle.
Because the SM throat is compactified, we should think of the dual
gauge theory as being coupled to gravity with a finite cut-off
$M_P$, rather than living on a fixed background spacetime.  In
nondynamical gravity, we would take $M_P\to\infty$ and renormalize
away divergent terms in the effective action. In this process, the
renormalized effective action would indeed yield a stress tensor
(and energy density) $\rho\propto q^2 H^4$. However, since we have
a finite Planck scale cut-off, no terms in the effective action
diverge.  Following along the calculation of
\cite{Christensen:1978yd}, we find that the energy density is
suppressed by powers of $H/M_P$ compared to (\ref{densitybh2}), so
the black brane does not actually match the dual gauge theory
during inflation.  In other words, not only does the black brane
fail to match the correct equation of state, but it also has wrong
magnitude of energy density.

\subsection{AdS-like Solutions}\label{sss:bucheltseytlin}

Another alternative is also inspired by string-gauge theory
duality;
\cite{Buchel:2001iu,Buchel:2002wf,Buchel:2002kj,Buchel:2003qm}
have found supergravity backgrounds dual to gauge theories on some
curved (but specified) spacetimes, including de Sitter spacetime.
Their supergravity duals are weakly curved everywhere and have
been used to study the motion of branes during inflation
\cite{Buchel:2003qj,Buchel:2004qg}, so we might expect that these
geometries could describe the SM throat during inflation. A
central observation in these constructions is that $AdS_5$ admits
a parameterization where the radial slices have a $dS_4$ geometry
\cite{Emparan:1999pm}.

The reason we do not think that the solutions of
\cite{Buchel:2001iu,Buchel:2002wf,Buchel:2002kj,Buchel:2003qm}
will describe the SM throat during inflation is somewhat
technical.  For specificity, let us focus on the solutions in
\cite{Buchel:2001iu,Buchel:2002wf}, which are appropriate for the
deformed conifold throats in our discussion.  These solutions are
found as perturbations around the de Sitter-like slicing of
$AdS_5\times T^{1,1}$, and, in fact, the perturbations disappear
at the bottom of the throat, leaving an $AdS_5\times T^{1,1}$
core.  Most importantly for us, the warp factor $A$ always
diverges at large distances from the core.  On the other hand, to
glue the throat onto a compact manifold, it should be possible to
find a solution with $A\to 0$ asymptotically (which is possible
for the flat spacetime solutions
\cite{Klebanov:2000nc,Klebanov:2000hb}). Suppose that we tried to
modify to solutions of \cite{Buchel:2001iu,Buchel:2002wf} to have
$A\to 0$ asymptotically.  Then asymptotically the solution would
go to $dS_4$ times a Ricci flat manifold.  However, the
supergravity stress tensor (sourced by supergravity fluxes) would
vanish asymptotically, leading to a contradiction.  This analysis
is born out by examining the equations of motion acting on a
supergravity ansatz that generalizes
\cite{Buchel:2001iu,Buchel:2002wf}\footnote{We thank A. Buchel for
conversations about this point.}.  However, it is possible that
the solutions of \cite{Buchel:2001iu,Buchel:2002wf} are
approximate solutions for throats in the compactification in a
region far from both the tip and the bulk.

This difficulty seems tied to the fact that these supergravity
solutions are dual to gauge theories on fixed spacetimes.  That is,
the gauge theories live in gravity backgrounds with infinite Planck
constant.  The Hubble scale of inflation therefore must be imposed by
hand.  In fact, because these dualities are derived by perturbation
around conformal theories, the Hubble scale is not an independent
physical mass scale in the problem; rather, it enters as a choice of
slicing for $AdS_5$.

In closing, we should also note that, were these solutions to reflect
the true SM throat geometry during inflation, the consequences for 4D
effective field theory would be disastrous.  As we mentioned above,
the core of these throats are $AdS_5\times T^{1,1}$.  In particular,
the warp factor $A\to -\infty$, leading to a horizon in the 10D
geometry and a vanishing effective string tension.

\section{More Reheating}\label{s:otherreheat}

Here we will discuss aspects of reheating in more general inflationary
models.

\subsection{Many Throats}\label{sss:manythroats}

The reader might wish to consider what happens if there exist
other throats with even more significant warping (in the true
ground state geometry) --- see related discussion in
\cite{Kofman:2005yz,Chialva:2005zy}.  
In the mechanism of \cite{Kachru:2003aw},
antibranes in these throats have small tensions, leading to a fine
spacing of possible values for the cosmological constant. Although
we are not concerned with the cosmological constant in this paper,
we should mention the effects of extra throats on reheating.

During the reheating phase, we expect that all the long throats
(that is, all the throats with ground state string scale less than
$H_\star$) are lifted to have $\ell_s H_\star \lesssim 1$.  The
argument is just the same as for the SM throat.  Therefore, when
the inflaton reheats the inflationary throat, the KK gravitons
should distribute themselves evenly among all the long throats. We
expect then that the energy density in the SM throat should be
reduced from our original considerations by a factor of the number
of long throats.\footnote{If these throats are distinguished by,
e.g., having different curvatures, the tunnelling to the
additional throats may be suppressed (or enhanced)
\cite{Langfelder}.} This effect would not seem to cause any
problems with reheating, since the number of long throats is
expected to be $\mathcal{O}(10)$.

Problems could arise, however, in low energy cosmology, such as
nucleosynthesis, if the energies in all the long throats remain
comparable.  Some mechanism is necessary to ensure that the SM
throat energy density dominates by late times.  An additional
complication is the fact that closed string modes (such as KK
gravitons) will prefer longer throats to the SM throat after about
$t\sim 1/\msm$, when the SM throat approaches its ground state
geometry.  The closed string modes will tunnel to the longer
throats, just as they tunnelled from the inflationary throat.
Fortunately, open string modes attached to the SM branes should
have a highly suppressed tunnelling rate and be effectively
``locked'' to the SM branes.  Perhaps this fact can account for
the dominance of the SM throat: if the SM throat has significantly
more branes than the other throats, then closed strings might
preferentially attach to the SM branes.  This will reduce the
closed string mode energy in the SM throat compared to the other
throats during an era in which all the long throats have similar
warp factors.  A subsequent redistribution of closed string energy
among the throats would then increase the total energy in the SM
throat.  We leave a further investigation for future work.

We can, however, question whether throats with such small warp factors
are phenomenologically viable in the present.  Such significant
warping would lead to very light KK gravitons, so these long throats
could be ruled out.  On the other hand, the light KK gravitons should
have highly suppressed wavefunctions in the SM throat, so SM modes
would couple to them very weakly.  It would be interesting to
determine whether such long throats are ruled out or not.

\subsection{Reheating from Other Inflationary Models}\label{sss:racereheat}

We can also give a few comments about reheating for other string
embeddings of inflation.

To put it shortly, both D3/D7 and non-BPS brane inflation should
reheat the SM throat similarly to the usual brane inflation case.
Both would generate closed strings in the inflationary region during
reheating, and the resulting KK modes could tunnel to the SM throat.

Now consider the case of racetrack inflation.  We recall first that
racetrack inflation is caused by the slow roll of a compactification
modulus, such as the total volume modulus of the compact manifold.
This modulus will generally couple both to SM brane modes and bulk
supergravity modes, so it can preheat the SM directly by parametric
resonance along with exciting closed string states.  Additionally, as
in brane inflation, the bulk supergravity modes can then reheat the SM
modes.


\section{More on String Thermodynamics}\label{s:morethermo}

In this appendix, we gather detailed formulae and calculations
relevant to the discussion of string thermodynamics in section
\ref{ss:thermo}.

\subsection{Open String Thermodynamics}\label{ss:open}

Here we collect some other useful formulae for open string
thermodynamics.  We work in the context of D3-branes only for
simplicity.

From (\ref{opensoup}), we can also see that the number density of open
strings is
\be
\label{avgrelate}
n_{o}=\frac{2aN_D^2}{b V_\perp}L\, .
\ee
Note that the open string system is extrinsic in the (noncompact)
directions parallel to the D3-branes.

We can also calculate the entropy from the string distribution
(\ref{opensoup})as in \cite{Lowe:1995nm,Lee:1997iz}.  The density of
states (per volume) for single open strings is
\be
\label{openstatedensity}
\omega_o(\varepsilon) = \frac{2aN_D^2}{b V_\perp}
e^{\beta_H\varepsilon}\ .
\ee
The density of states for a gas of strings of energy $E$ in a volume
$V_{\|}$ along the branes is therefore
\be\label{opengasstatedensity}
\Omega_o =\left(\frac{C}{E}\right)^{1/2} e^{\beta_H E} I_1\left(
2\sqrt{CE}\right)\propto \exp\left[ \sqrt{\frac{8a N_D^2 V_{\|} E}{b
\mu V_\perp}}+\beta_H E\right]\ ,
\ee
where $4CE$ is the combination inside the square root after the
proportionality and $I_1$ is a modified Bessel function
\cite{Abel:1999rq}.  The proportionality is valid in the limit that
$E\to\infty$, $E/V_{\|}$ constant.  Therefore, the entropy is
\be\label{opengasentropy} S_o \approx \beta_H E+ \sqrt{\frac{8a N_D^2
V_{\|} E}{b \mu V_\perp}}\ .\ee
It is also possible to argue that quantum corrections to the entropy
are no more important than the low energy string states in the
Hagedorn limit, though we do not give the details here.

\subsection{Closed String Thermodynamics}\label{ss:closed}

Here we give a very brief review of closed string thermodynamics (in
contrast to open strings), which is relevant for the behavior of
cosmic strings.  The density of states was given in \cite{Deo:1989bv};
a recent review (with more references) is \cite{Barbon:2004dd}, and
the relevance to cosmic strings is reviewed in \cite{Vilenkin}.

In the relevant case of 3 noncompact spatial dimensions, the density
of states per volume for a single closed string is
\be\label{closed1}
\omega_c(\varepsilon) \approx \frac{1}{\ap{}^3 \varepsilon^{5/2}}
e^{\beta_H \varepsilon}\ .\ee
The density of states for the entire gas of strings is
\be\label{closed2}
\Omega_c\propto \frac{V}{(E-V/\ap{}^2)^{5/2}} \exp\left[\beta_H E+\ap{}^{-3/2}
V\right]\ .\ee
This density of states leads to a distribution of closed strings with
one (or a few) very long, energetic strings, and a gas of closed
string radiation with total energy density $\rho\sim \ap{}^{-2}$.

Additionally, it has been argued that D-branes are very efficient at
chopping closed strings into open strings.  That is, if the D-branes
are packed with a density about unity in string units in the space
transverse to their worldvolumes, open strings will dominate closed
strings in equilibrium.

\subsection{Hagedorn Strings vs Black Branes}\label{ss:vsbbrane}

We will argue here that a Hagedorn gas of open strings will
\textit{not} undergo a phase transition to a black 3-brane, as is
commonly supposed in the literature (and does indeed happen in some
cases) (see \cite{Horowitz:1997nw,Barbon:2004dd}).  For simplicity, we
will consider the case of D3-branes only.  We will also work in flat
space as an approximation.

Let us start by considering the open string gas and black 3-brane
in the microcanonical ensemble.  In the open string entropy
(\ref{opengasentropy}), the ratio of the first to second terms is
$(\hat\beta/2\pi B)(\rho/N_D^2\mu^2)^{1/2}$, where we have defined
$\beta_H=\hat\beta\sqrt{\ap}$ and $B=8a\ap/b\mu V_\perp$ (both
$\hat\beta$ and $B$ are $\mathcal{O}(1)$).  Therefore, for
densities larger than $N_D^2\mu^2$ --- where the average string is
longer than the string length --- the first term in
(\ref{opengasentropy}) dominates and the entropy is $S_o\approx
\beta_H E$. On the other hand, the entropy of the black 3-brane is
$S_{bb}= A\sqrt{N_D} V^{1/4} E^{3/4}$ \cite{Gubser:1996de}, where
$A$ is a constant of order unity. Then the ratio of entropies is
\be\label{sratio1}
\frac{S_o}{S_{bb}} \approx \frac{\hat\beta}{\sqrt{2\pi}A}\left(
\frac{\rho}{N_D^2\mu^2}\right)^{1/4}\gg 1\ee
for Hagedorn densities.  Heuristically, this result makes sense;
starting at low densities, we expect a radiation bath to undergo a
phase transition to a gas of long open strings.  However, the
radiation bath has the same entropy as the black brane
\cite{Gubser:1996de}, so we would not expect a further phase
transition to a black brane phase.

In the canonical ensemble, some manipulations give the free energies
\be\label{fe}
F_o = -\frac{1}{4} \frac{B^2 N_D^2 V}{\ap \beta(\beta-\beta_H)}\ ,\ \
F_{bb}= -\frac{27}{256} \frac{A^4 N_D^2 V}{\beta^4}
\ee
for temperatures below the Hagedorn temperature $1/\beta_H$.  The
ratio of free energies is
\be\label{feratio}
\frac{F_o}{F_{bb}} = \frac{64}{27} \frac{B^2}{\ap A^4}
\frac{\beta^3}{\beta-\beta_H} = \frac{64}{27} \frac{B\hat \beta^3}{\pi A^4}
\left(\frac{\rho}{N_D^2\mu^2}\right)^{1/2} \ .\ee
For small enough $\beta$ (or high enough density), the open string
free energy is larger (i.e., more negative) and so the open
strings are thermodynamically favored.  We should contrast this
behavior to that expected from the AdS/CFT correspondence, in
which a thermal state of the gauge theory corresponds to a black
brane horizon in the throat for high temperatures.  In that case,
the strings in the throat are exposed to a heat bath of a
temperature greater than the Hagedorn temperature.  Therefore, the
only thermal state available to any open strings in the throat is
a black brane. Furthermore, we expect that closed strings will
actually have a phase transition to a black brane, so there is no
issue for gravity duals with no free branes.

There are some possible loopholes in our argument, however.  For
example, we have neglected any effects from the nontrivial geometry of
the SM throat, instead using a flat background.  In particular, the
changing warp factor may favor a black brane horizon.  If the strings
filled the entire throat and entered the region of the
compactification where $A\sim 0$, then they should be compared to a
far-from-extremal black brane, which has a different entropy formula.
We leave it to the future to check our results.

\subsection{Worldsheet Fluctuations in Warp Factor}\label{ss:wsfluct}

In this appendix, we will discuss the compact volume $V_\perp$
accessible to long strings in the SM throat.  Classically speaking,
strings are confined to minima of the warp factor because the strings
feel $e^{2A}$ as a gravitational potential.  However, in a quantum
mechanical or thermal system, the string centers of mass can spread to
larger values of the warp factor, as was first discussed in
\cite{Jackson:2004zg}.  We will first review the zero temperature
calculation of \cite{Jackson:2004zg}, and then we will generalize it
to finite temperature.  In the end, we will find that the two
calculations give similar answers.

In the metric (\ref{warpmetric}), the static gauge action for
fluctuations of a long string in the compact dimension $Y$ is
\be\label{staticS}
S = -\frac{1}{2\pi\ap}\int d^2\sigma \left[e^{2A} +\frac{1}{2}\left(
\del_a Y\del^a Y+ (\del_Y^2 e^{2A})Y^2\right)+\cdots\right]\ .\ee
Here, we are using diffeomorphism invariance in the internal
dimensions to simultaneously diagonalize $g_{mn}$ and
$\del_m\del_n e^{2A}$. The $\cdots$ represent interactions due to
higher derivatives of the warp factor; note that the mass is given
by $m^2=\del_Y^2 e^{2A}$.  The deviation of the string in
direction $Y$ is just given by the propagator $\langle Y^2\rangle$
on the Euclidean worldsheet.  Then we have
\be\label{quantum1} \langle Y^2\rangle =
\frac{\ap}{2\pi}\int^\Lambda \frac{d^2k}{k^2+m^2} =\frac{\ap}{2}
\ln\left[ 1+\frac{\Lambda^2}{m^2}\right]\ ,
\ee
where $\Lambda$ is a worldsheet UV cutoff.  The natural cutoff
scale for the worldsheet is the string scale measured by the 4D
coordinate time (which is the worldsheet time in static gauge)
because the string scale is the gap for excitations of string in
the noncompact directions.  When studying the conifold throats,
then, \cite{Jackson:2004zg} had $\Lambda^2=e^{2A}/\ap$ and
$m^2=e^{2A}/g_s n_R\ap$.  Further, at the tip of the conifold
throat, the topology of the transverse geometry is
$\mathbf{R}^3\times S^3$, where the warp factor varies only in the
$\mathbf{R}^3$ directions. Therefore, the strings can fill the
$S^3$ at the tip as well as climb slightly up the warp factor
potential well.  To calculate the effective volume in the
$\mathbf{R}^3$, we use $\sqrt{\langle Y^2\rangle}$ as the linear
scale.  In the end, we arrive at \cite{Jackson:2004zg}
\be\label{jjpvol} \frac{V_\perp}{(4\pi^2\ap)^3} \approx \frac{(g_s
n_R)^{3/2}}{4\pi} \left[ \frac{\ln (1+g_s n_R)}{2\pi}\right]^{3/2}\ .
\ee
For $g_s n_R\sim 10$, $V_\perp/(4\pi^2\ap)^3\sim 1/2$.  We should
regard this volume as a minimum value, regardless of whether the gas
of strings can stay in thermal equilibrium.

Near the Hagedorn temperature, though, we should use a thermal
expectation value rather than the quantum value.  Now the Euclidean
time direction is compact, and the two-point function is
\be\label{thermo1}
\langle Y^2\rangle = \frac{\ap}{\beta} \sum_{n=-\infty}^\infty
\int dk \frac{1}{\omega_n^2+k^2+m^2}\ ,\ \ \omega_n = \frac{2\pi n}{\beta}\ ,
\ n\in\mathbf{Z}\ .\ee
The momentum $k$ is now only in the spatial worldsheet direction.
After the (convergent) momentum integral,
\be\label{thermo2}
\langle Y^2\rangle = \frac{\pi\ap}{\beta}\sum_n
\frac{1}{(\omega_n^2+m^2)^{1/2}}\ .
\ee
The sum diverges, so once again we need to impose the same cutoff
$\Lambda$ for the frequency.  Further, we are interested in the
Hagedorn limit, so we use $\beta=\beta_H\sim 2\pi\lst$.  Therefore,
only the $n=0$ term contributes, due to the cutoff, and using the same
mass as before, we get
\be\label{thermo3}
\langle Y^2\rangle \approx \frac{\ap}{2} \sqrt{g_s n_R}\ .\ee
The volume is then
\be\label{thermalvol} \frac{V_\perp}{(4\pi^2\ap)^3} \approx \frac{(g_s
n_R)^{3/2}}{4\pi} \left[ \frac{\sqrt{g_s n_R}}{2\pi}\right]^{3/2}
\ee
in a thermal calculation.  Plugging in the numerical values from
above, we see $V_\perp/(4\pi^2\ap)^3\sim 1$.  Since there is not much
difference between the quantum and thermal values, we can just
approximate $V_\perp/(4\pi^2\ap)^3\sim 1$ whenever we need to make an
estimate.

We conclude with one additional note.  Above, we implicitly
treated the string as being long by integrating over spatial
worldsheet momenta.  For short strings (those near $\lst$ in
length), we should replace the spatial momentum integral by a sum.

\bibliography{reheat}

\end{document}